\documentclass[epj,nopacs]{svjour}

\usepackage{amssymb}
\usepackage{epsfig}
\usepackage{eucal}
\usepackage{amsmath}

\newcommand{\beq}{\begin{equation}}
\newcommand{\eeq}{\end{equation}}
\newcommand{\bea}{\begin{eqnarray}}
\newcommand{\eea}{\end{eqnarray}}
\newcommand{\non}{\nonumber}

\headnote{\vspace{-.5cm}
\hfill \parbox{4cm}{\vspace{-2cm}
\tt \normalsize CERN-TH/2002-328 \\ TTP02-43}}

\title{The radiative return at \(\mathbf  \phi\)- and B-factories:
 small-angle photon emission at next to leading order
\thanks{Work supported in part by BMBF under grant number 05HT9VKB0, 
 EC 5-th Framework EURIDICE network  project HPRN-CT2002-00311,
 EC 5-th Framework contract HPRN-CT-2000-00149,
 TARI project HPRI-CT-1999-00088, 
 Polish State Committee for Scientific Research (KBN) under 
 contract No. 2 P03B 017 24, MCyT Plan Nacional I+D+I (Spain)
 under grant BFM2002-00568 and Generalitat Valenciana 
 under grant CTIDIB/2002/24.}}

\author{Henryk Czy\.z\inst{1,2}\thanks{\email{czyz@us.edu.pl}} 
\and Agnieszka Grzeli{\'n}ska\inst{1,2}\thanks{Supported in part 
by `Marie Curie Training Site' at Karlsruhe Univ.;
\email{grzel@joy.phys.us.edu.pl}}
\and Johann H. K\"uhn\inst{1}\thanks{\email{jk@particle.uni-karlsruhe.de}}
\and Germ\'an Rodrigo\inst{3}\thanks{Supported in part by
E.U. TMR grant HPMF-CT-2000-00989; \email{german.rodrigo@cern.ch}}
} 

\institute{
Institut f\"ur Theoretische Teilchenphysik,
Universit\"at Karlsruhe, D-76128 Karlsruhe, Germany. \and
Institute of Physics, University of Silesia,
PL-40007 Katowice, Poland. \and
Theory Division, CERN, CH-1211 Geneva 23, Switzerland.}

\date{Received: December 16, 2002}

\abstract{ The radiative return offers the 
 unique possibility for a measurement 
 of the cross section of electron--positron annihilation into hadrons
 over a wide range of energies. The large luminosity of present
 \(\phi\)- and B-factories easily compensates for the additional factor
 of \(\alpha\) due to the emission of a hard photon. Final states with photons
 at large angles can be easily identified. The rate for events with
 collinear photons, however, is enhanced by a large logarithm and allows,
 in particular at lower energies, for a complementary measurement.
 The Monte Carlo generator PHOKHARA, which includes next to leading order
 corrections from virtual and real photon emission, has been extended
 from large photon angles into the collinear region, using recent
 results for the virtual corrections. In addition, the present version
 includes final state radiation for muon and pion pair production and
 final states with four pions. Implications for the experimental 
 analysis at three typical energies, 1.02, 4 and 10.6 GeV, are presented:
 the magnitude of these new corrections is studied, possibilities
 for the separation of initial and final state radiation are proposed,
 and the differences with respect to the previous treatment based on structure 
 functions are investigated.
  }

\begin{document}

\def\Li{\hbox{Li}}                                            
 

\maketitle

\section{Introduction}

 Measurements of the cross section for electron--positron annihilation into 
 hadrons are essential for the interpretation of the recent, precise results
 for the muon anomalous magnetic moment \(a_\mu\) \cite{Brown:2001mg}.
 Similarly they are relevant for our knowledge of
 the running of the fine structure constant
 and thus crucial for the analysis of electroweak precision
 measurements at high energy colliders \cite{Jegetc.,HMNT02,Davier:2002dy}. 

Of particular importance for these two applications is the low energy region,
say from threshold up to centre-of-mass system (cms) energies of approximately 
3~GeV for \(a_\mu\) and 10~GeV for \(\alpha(M_Z)\).
 Recent measurements based on energy scans 
between 2 and 5~GeV have improved the accuracy in part of this
range \cite{BES}. Similar, or even further improvements below 2~GeV
would be highly welcome. The region between 1.4~GeV and 2~GeV, in
particular, is poorly studied and no collider will cover this region 
in the near future. Improvements  
on the precise measurements of the pion form factor in the low 
energy region by the CMD2 and DM2 collaborations \cite{CMD2},
or even an independent cross-check,
would be extremely useful, in particular in view of the 
disagreement between \(e^+e^-\)data and the analysis 
based on isospin-breaking-corrected 
$\tau$ decays~\cite{Davier:2002dy}.

 Traditionally the
 energy dependence of the cross section was deduced from experiments,
 where the beam energy was varied over the range dictated by the energy
 reach of the collider. This `energy scan' allows, at a first glance,
 a fairly simple interpretation of the measurement in terms of the so called 
 R-ratio, which enters the aforementioned applications. Nevertheless also
 in this case initial state radiative corrections (ISR) give rise 
 to complications and require a complicated unfolding procedure discussed
 below.

 As an alternative the `radiative return' has been suggested 
 \cite{Binner:1999bt,Melnikov:2000gs,Czyz:2000wh,Spagnolo:1999mt} as 
 a particularly attractive option for \(\phi\)- and B-meson factories.
 These collider experiments operate at fixed energies, albeit with enormous
 luminosities. BABAR and BELLE at 10.6 GeV, CLEO-C in the region
 between 3 and 5 GeV and KLOE at 1.02 GeV are the experiments of interest
 for the subsequent considerations.
 This peculiar feature of a `factory' allows the use of the radiative return, 
 i.e. the reaction
\bea
e^+(p_1)+e^-(p_2) \to \gamma(k_1) + \gamma^*(Q)\left(\to {\rm hadrons}\right)
 \ \ ,
\label{eq1}
\eea
to explore a wide range of \(Q^2\) in a single experiment.

Nominally invariant masses of the hadronic system between 
the production threshold of the respective channel
 and the cms energy  
of the experiment are accessible.
In practice, to clearly identify the reaction, it is useful to consider 
only events with a hard photon--tagged or untagged--which lowers the
 energy significantly.

To arrive at reliable predictions for differential and for partially 
integrated cross sections, including kinematical cuts as used
 in experiments, a Monte Carlo generator is indispensable.
 The inclusion of radiative corrections in the generator and 
 in the experimental analysis
 is essential for the precise extraction of the cross section. For hadronic
 states with invariant masses below 2 to 3 GeV, it is desirable
 to simulate the individual channels with two, three and up to six mesons,
 i.e. pions, kaons, \(\eta\)'s, etc., which requires a fairly detailed 
 parametrisation of various form factors.

 A first program called EVA was constructed some time ago \cite{Binner:1999bt}
 to simulate the production of a pion pair together with a hard photon.
 It includes initial state radiation, final state radiation (FSR), their 
 interference, and the dominant radiative corrections from additional
 collinear radiation through structure function (SF) technique \cite{CCR}.
 A similar program that simulates the production of four pions 
 together with a hard photon has been developed in \cite{Czyz:2000wh}.
 More recently a new Monte Carlo generator called 
 PHOKHARA \cite{RCKS}
 was developed. It includes, in contrast to the former generators, the
 complete next-to-leading order (NLO) radiative corrections. 

The first version
 of PHOKHARA incorporates ISR only and is limited to 
 \(\pi^+\pi^-\gamma(\gamma)\) and \(\mu^+\mu^-\gamma(\gamma)\) 
 as final states. PHOKHARA exhibits, however, 
 a modular structure that simplifies the
 implementation of additional ha-dronic modes or the replacement of the 
 current(s) of the existing modes. Its first version was designed
 to simulate  
 configurations with photons emitted at relatively large angles,
 \(\theta^2 \gg m_e^2/s \). In this case it is legitimate to drop terms
 proportional to \(m^2_e\), an assumption that leads to a considerable
simplification of the virtual corrections \cite{Rodrigo:2001jr}. Subsequently
 analytical results for the virtual corrections, that
  are also valid into the small angle region, 
  were obtained in \cite{RK02}. 
 The extension of the program PHOKHARA into this small angle region,
 incorporating these new analytic results are the central topic 
 of the present paper. The description of this new feature 
 and numerous tests of the program 
 stability and technical precision are contained in Section 2.

 Final state radiation can affect 
 the measurement of the pion form factor, and quite generally of the R-ratio.
 However, using suitable cuts, its effects can be significantly reduced. 
 Moreover, given sufficiently large event rates its magnitude can be
 extracted experimentally by varying the cuts and/or comparing events
 with different photon angles with respect to beam and pion directions,
 respectively. The charge asymmetry that arises from ISR--FSR interference
 provides another important handle on this `background'. For this reason
 FSR from \(\pi^+\pi^-\) and \(\mu^+\mu^-\) has been incorporated in 
 the upgrade of the program and will be discussed in Section 3. 
 The \(\mu^+\mu^-\) final state is still limited to its QED part,
 e.g. the narrow resonances (J/\(\psi\)) are not (yet) included.

 The program has also been extended to include final states 
 with four pions, following the lines discussed  in \cite{Czyz:2000wh}.
 The implementation of these new channels will be discussed in Section 4.

\section{The radiative return for small-angle emission and tests 
 of the program} 

 The study of events with photons emitted under both large and small angles,
 and thus at a significantly enhanced rate, is particularly attractive for the
 \(\pi^+\pi^-\) final state with its clear 
 signature~\cite{Aloisio:2001xq,Denig:2001ra,Adinolfi:2000fv,Barbara:Morion,Venanzoni:2002,Achim:radcor02}.
 In contrast events with a tagged photon, emitted at a large angle, have a 
 clear signature particularly suited to the analysis of hadronic
 states of higher multiplicities \cite{babar,Berger:2002mg}.

 The inclusion of radiative corrections is essential for the precise extraction
 of the cross section, which is necessarily based on a Monte Carlo simulation.
The complete NLO corrections have recently been implemented in 
the program PHOKHARA. However, just like the earlier EVA, 
this program was designed for photon emission at large angles 
(`tagged photons'). For nearly collinear photons,
 corrections from virtual and real photon emission,
 as well as Born terms, must include those contributions proportional to
\(m_{e}^{2}/s\) and even to \(m_{e}^{4}/s^{2}\), which are enhanced by their 
highly singular angular dependence and thus integrate to terms
of order 1 and proportional to  \(\alpha/\pi\)
for Born and  NLO terms respectively. These
mass-suppressed terms are significantly smaller than the leading,
logarithmically enhanced pieces; they must nevertheless be taken into
account for a consistent treatment. The evaluation of corrections from 
virtual plus soft photon emission to reaction (\ref{eq1}), valid for the
full angular region, has been treated in \cite{RK02}. Essentially it
consists of the calculation of the leptonic tensor \(L_{\mu\nu}\), which 
has to be multiplied by the hadronic tensor \(H_{\mu\nu}\), so that
a fully differential distribution is obtained. To arrive at a reasonably
compact, numerically stable result, the limit \(m_{e}^{2}/s\ll1\) 
for \(L_{\mu\nu}\) is considered. However, terms proportional to \(m_{e}^{2}\)
must be kept if these exhibit the singular angular dependence discussed above.

\begin{figure*}
\begin{center}
\epsfig{file=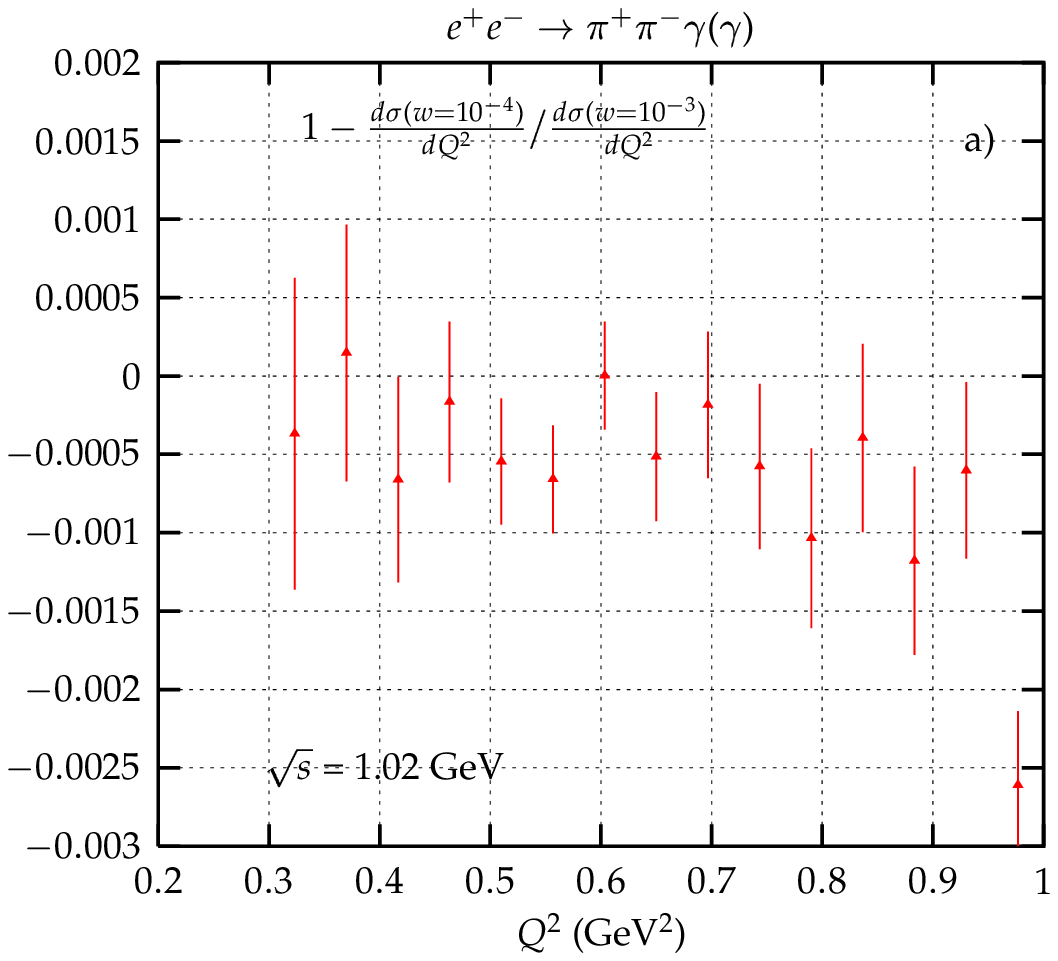,width=7.5cm}\hskip 2 cm
\epsfig{file=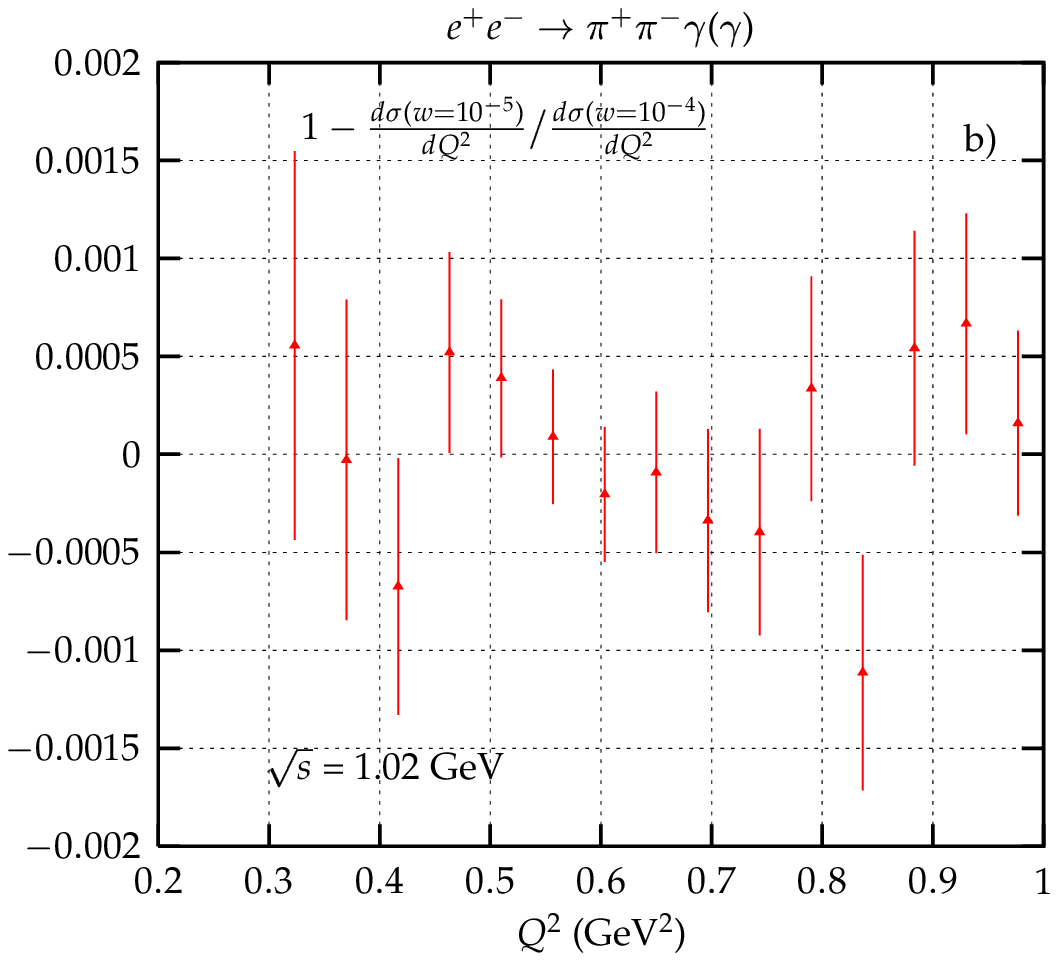,width=7.5cm}
\caption{Comparison of the $Q^2$ differential distribution for different
values of the soft photon cutoff: $w = 10^{-3}$ vs. $10^{-4}$ and
$w = 10^{-4}$ vs. $10^{-5}$ ,
at $\sqrt{s}=1.02$~GeV. One of the photons was required 
to have energy $>$ 10 MeV. No further cuts were applied.}
\label{fig:eps1}
\end{center}
\end{figure*} 

\begin{figure*}
\begin{center}
\epsfig{file=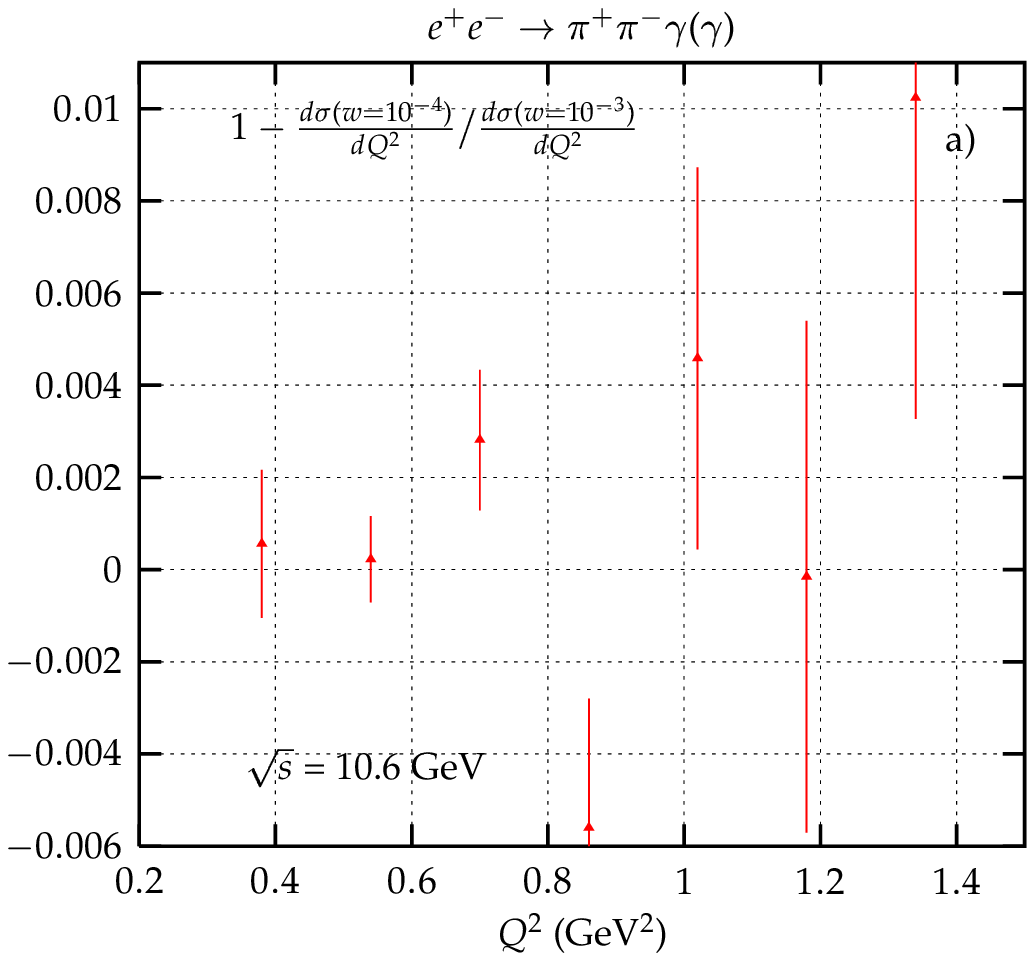,width=7.cm}\hskip 2 cm
\epsfig{file=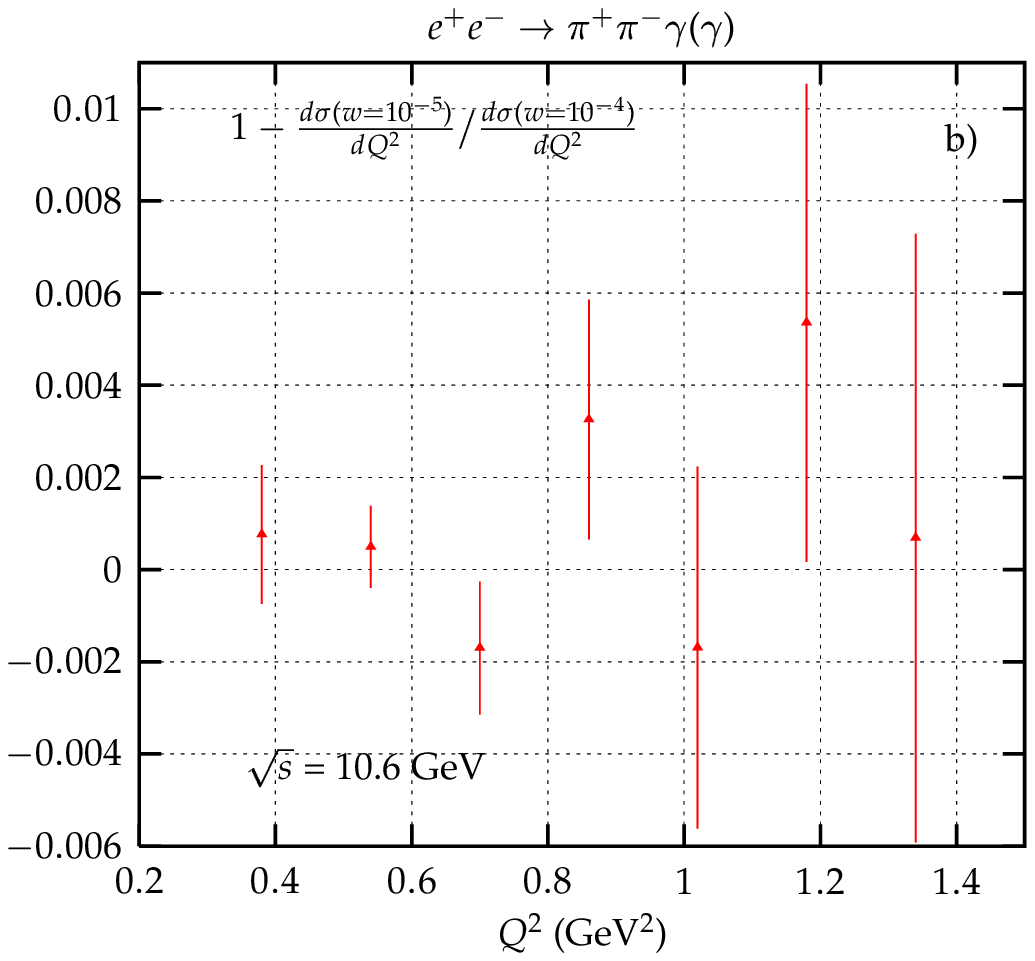,width=7.cm}
\caption{Comparison of the $Q^2$ differential distribution for different
values of the soft photon cutoff: $w = 10^{-3}$ vs. $10^{-4}$ and
 $w = 10^{-4}$ vs. $10^{-5}$ ,
at $\sqrt{s}=10.6$~GeV. One of the photons was required 
to have energy $>$ 100 MeV.
 No further cuts were applied.}
\label{fig:eps3}
\end{center}
\end{figure*}

The differential rate for the virtual and soft QED corrections is thus
cast into the product of a leptonic and a hadronic tensor 
 and the corresponding factorised phase space:
\bea
d\sigma &=& \frac{1}{2s}L_{\mu\nu}d\Phi_2(p_1+p_2;Q,k_1) \nonumber \\
 &\times & H^{\mu\nu} d\Phi_n(Q;q_1,\cdots,q_n) \frac{dQ^2}{2\pi} \ \ ,
\label{eq2}
\eea

where \(d\Phi_n(Q;q_1,\cdots,q_n)\) denotes the hadronic \(n\)-body
 phase
space, including all statistical factors, \(Q^2\) is the invariant mass 
 of the hadronic system and \(d\Phi_2(p_1+p_2;Q,k_1)\) is 
 the two-body phase space of the photon and the hadronic system.
The tensor \(L_{\mu\nu}\) 
depends on the four-vectors \(p_{1}\), \(p_{2}\), \(Q\), \(k_{1}\) and
the soft photon cutoff \( w\equiv E_{\gamma}^\mathrm{max}/\sqrt{s}\). Its
explicit functional form is given in \cite{RK02}. 
The description of the hadronic system is model-dependent.
It enters only through the hadronic tensor
\bea
H^{\mu\nu} = J^\mu J^{\nu \dagger} \ ,
\label{eq3}
\eea
where the hadronic current has to be parametrised through form factors 
\cite{Czyz:2000wh,Kuhn:1990ad,Decker:1994af,Ecker:2002cw}. The running
 of \(\alpha\) is not taken into account in this program and can be included
 trivially in the final experimental analysis.

\begin{table}
\caption{Total cross section (nb) for the process
$e^+ e^- \rightarrow \pi^+ \pi^- \gamma$ at NLO for different values
of the soft photon cutoff. Only initial state radiation.
One of the photons with energy larger than 10 MeV for \(\sqrt{s} = 1.02\) GeV 
and larger than 100 MeV for \(\sqrt{s} = 10.6\) GeV. $Q^2<1$GeV.
No further cuts applied.}
\label{tab:epstest}
\begin{center}
\begin{tabular}{ccc}
$w$ & $\sqrt{s}=$1.02~GeV  & 10.6~GeV \\ \hline
$10^{-3}$ & 36.999 (3)  & 0.15557(7)\\
$10^{-4}$ & 37.021 (3)  & 0.15548(6)\\
$10^{-5}$ & 37.021 (3)  & 0.15545(6)\\ \hline
\end{tabular}
\end{center}
\end{table}                          
The matrix element for the emission from the initial state of two real 
hard photons, i.e. \(E_{\gamma_i} > w\sqrt{s}\), with \(i=1,2,\)
\bea
e^+(p_1)+e^-(p_2) \to  \gamma^*(Q) + \gamma(k_1) + \gamma(k_2)
 \ \ ,
\label{eq4}
\eea
is calculated numerically following the helicity amplitude method 
with the conventions introduced in \cite{Kolodziej:1991pk,Jegerlehner:2000wu}.
The results from \cite{RCKS}, which were used for tagged photon 
events are equally applicable for the present purpose.
 \begin{figure}[!ht]
\begin{center}
\epsfig{file=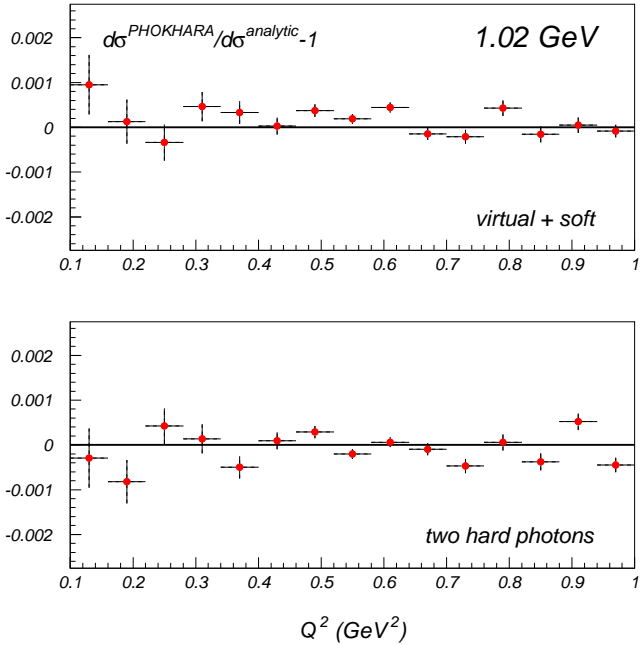,width=9.5cm} 
\end{center}
\caption{Comparison of the virtual+soft and hard contributions to the 
$\pi^+ \pi^-$ differential cross section with inclusive analytical results.
Soft photon cutoff: $w=10^{-4}$.}
\label{fig:realvirt}
\end{figure}
\begin{figure*}[!ht]
\begin{center}
\epsfig{file=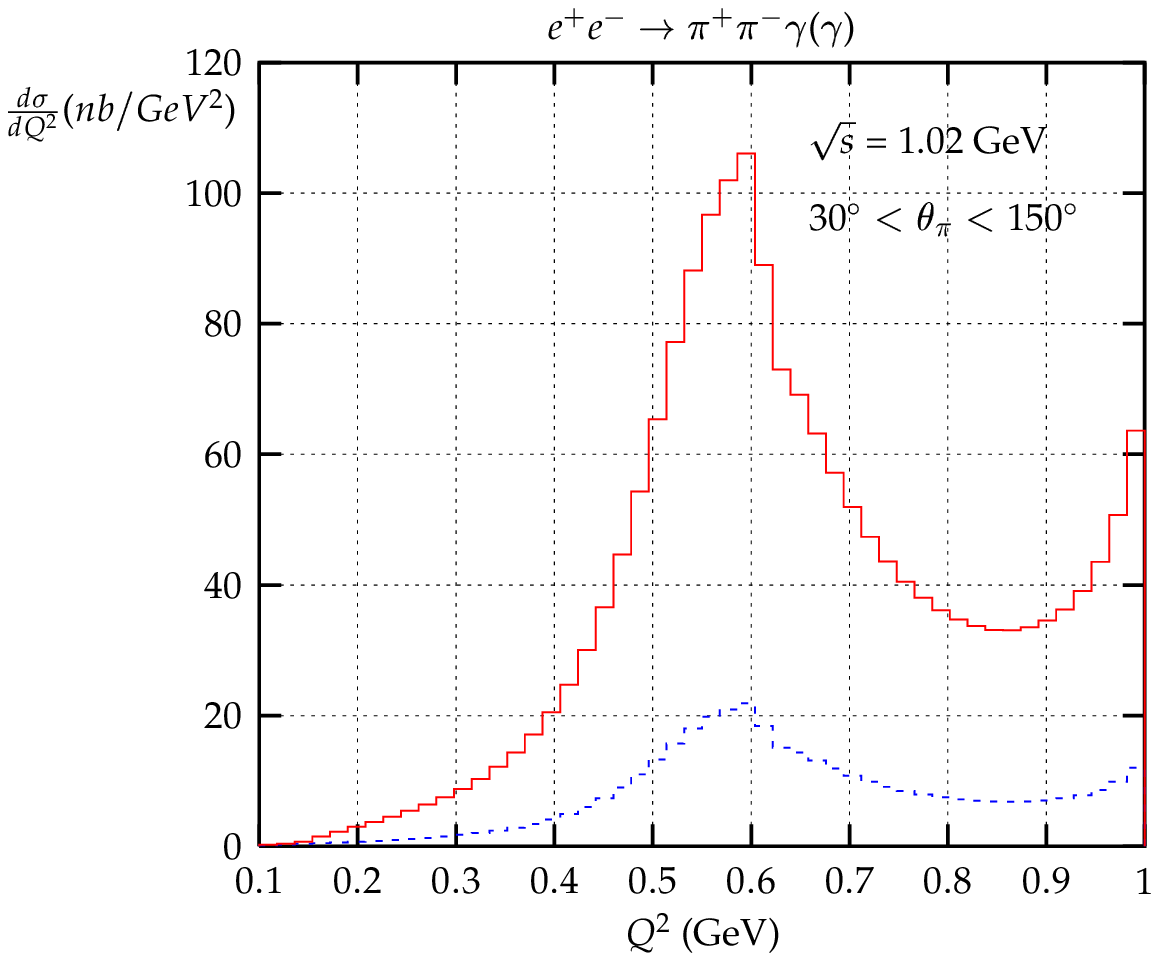,width=8cm}\hskip 1 cm
\epsfig{file=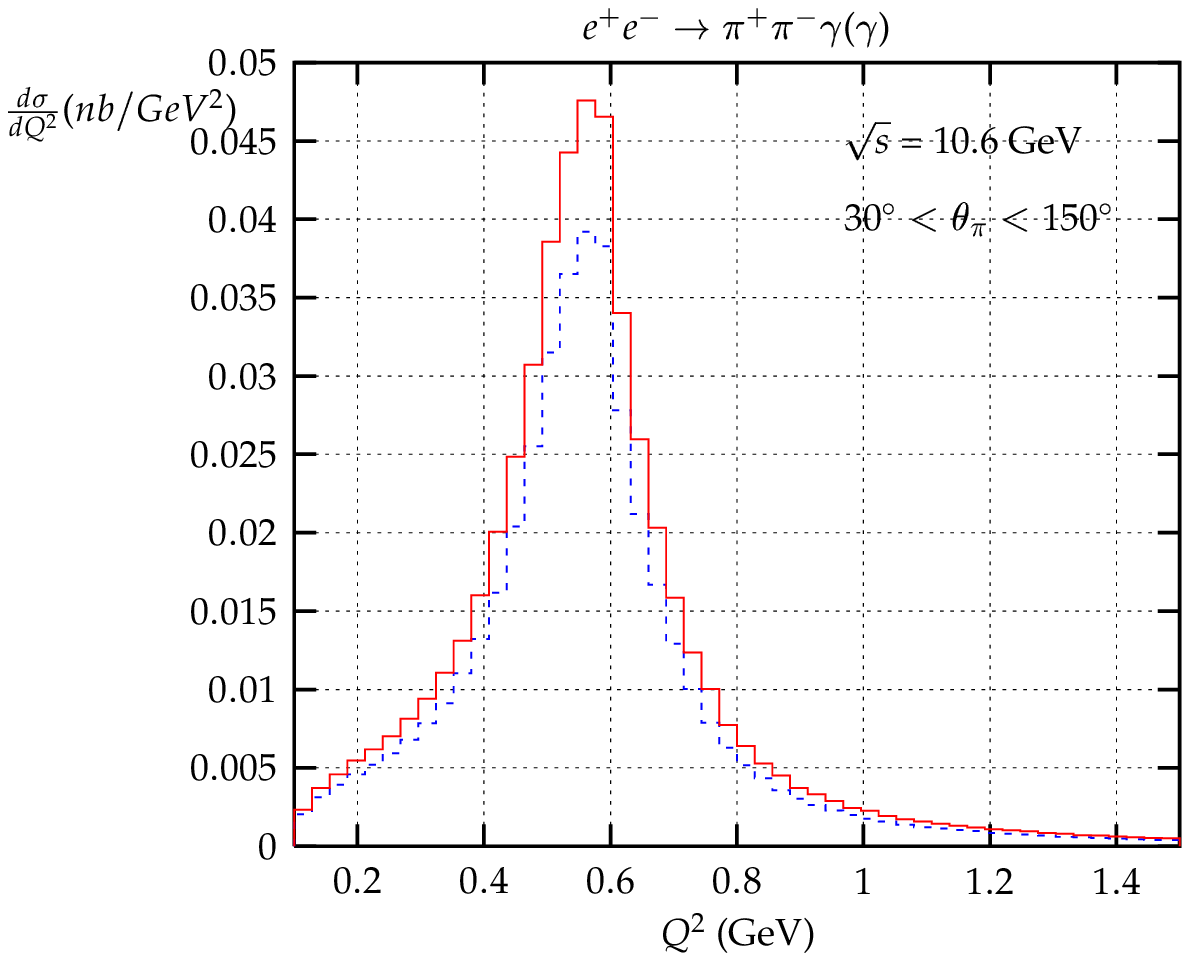,width=8cm}
\caption{Comparison of the $Q^2$ differential cross sections
 for \(\sqrt{s} =\) 1.02 (left) and 10.6 GeV (right).
  The pion angles are restricted to 
 \(30^\circ < \theta_{\pi^\pm } < 150^\circ\), while the photon(s)
 angles are not restricted (upper curves) and restricted to 
 \(30^\circ < \theta_\gamma < 150^\circ\) (lower curves).}
\label{fig:sig_cuts}
\end{center}
\end{figure*} 
                                                     
The virtual plus soft photon contribution and the hard one depend separately
on the soft photon cutoff \(w\) used to regulate the infrared divergences
of the virtual diagrams. The former shows a logarithmic \(w\)-dependence.
The latter, after numerical integration of the phase space, exhibits the
 same behaviour, whereas their sum must be independent of \(w\). To show
 that this indeed occurs is therefore a basic test of the performance 
 of the program. The value of \(w\) that optimises the efficiency of 
 the event generation,
 avoiding at the same time the appearance of the negative weights, is 
 determined by this procedure.

 Table~\ref{tab:epstest} presents the total cross section 
for radiative production of a pair of pions calculated 
for several values of the soft photon cutoff at two different cms 
energies. The energy of one of the photons was required to be larger
than 10 MeV for \(\sqrt{s} = 1.02\) GeV and larger than 
 100 MeV for \(\sqrt{s} = 10.6\) GeV.
No further kinematical cuts were applied, thus allowing  
to test in particular the small photon 
angle region.

For \(\sqrt{s}=1.02\) GeV the comparison 
between \(w = 10^{-3}\) and \(10^{-4}\) indicates a residual 
 \(w\)-dependence.
The excellent agreement between \(w = 10^{-4}\) and \(10^{-5}\), 
within the error of the numerical integration, 
confirms the $w$-independence of the result. 
A value around $w=10^{-4}$ seems to be the best choice
as observed before for large angle photons \cite{RCKS,Rodrigo:2002hk}.

In Figs. \ref{fig:eps1} and \ref{fig:eps3}
the \(Q^{2}\) dependence of the differential cross section \(d\sigma/dQ^{2}\)
is compared for different choices of the cutoff, after integration over 
the remaining kinematic variables. Again for \(\sqrt{s}=1.02\) GeV
 the comparison 
between \(w = 10^{-3}\) and \(10^{-4}\) shows a residual \(w\) 
dependence (Fig. \ref{fig:eps1}a), 
which disappears beyond \(w = 10^{-4}\) (Fig. \ref{fig:eps1}b).
At a cms energy of 10.6 GeV  the result is numerically stable for 
\(w = 10^{-3}\) already (Fig. \ref{fig:eps3}a). 
 Stable results are also obtained for \(w\) around and
below \(10^{-4}\) (Fig. \ref{fig:eps3}b). 
Thus \(w = 10^{-4}\) is used as the default value in 
the program. Similar tests were performed for the four-pion channels
 (see Section 4).

The present implementation of PHOKHARA covers the full angular
region for photon emission. This allows for a number of tests and 
comparisons with analytical results that were not possible with 
the previous version. In Fig. \ref{fig:realvirt} the results of the program 
are compared with the analytical results from Ref. 
 \cite{Berends:1988ab}. We use their Eqs. (2.25)+(2.26) for the virtual
 plus soft part and Eq. (2.28) for the hard emission part.
As it was necessary to change several couplings in the original
 formulae, we repeat (Appendix B) the expressions actually used
 for the comparison. 
 Agreement within the 
statistical uncertainty and in any case better than \(10^{-3}\) is 
evident from this comparison.

Initial state radiation is dominated by photons at small angles. Inclusion of
events with nearly collinear photons thus leads to a significant 
enhancement of the observed event rate. The comparison between 
the differential cross sections for large angle photon events 
\((30^{\circ}<\theta_{\gamma}<150^{\circ})\) and without restriction on 
\(\theta_{\gamma}\) is shown in Fig. \ref{fig:sig_cuts}.
 The pion angles are always
assumed to be restricted to the region 
\(30^{\circ}<\theta_{\pi^{\pm}}<150^{\circ}\). 
Results are presented for two different cms energies
 ( \(\sqrt{s} = \) 1.02 and 10.6 GeV). There is a big quantitative difference
 between these two energies. For \(\sqrt{s}\) = 1.02 GeV a huge contribution
 from small angle photons is observed for the full range of \(Q^2\).
 In contrast
  the gain in the cross section for  
 \( \sqrt{s} = \) 10.6 GeV is small
 as a consequence of the conflicting kinematical constraints of small photon
 and large pion angles.
 FSR has not been included in these figures. For 10.6 GeV its contribution 
 is negligible, while for 1.02 GeV and for the cuts used for 
 Fig. \ref{fig:sig_cuts} it is sizeable, but can be reduced using the cuts
 discussed below.

\begin{figure}
\vspace{-.8cm}
\begin{center}
\epsfig{file=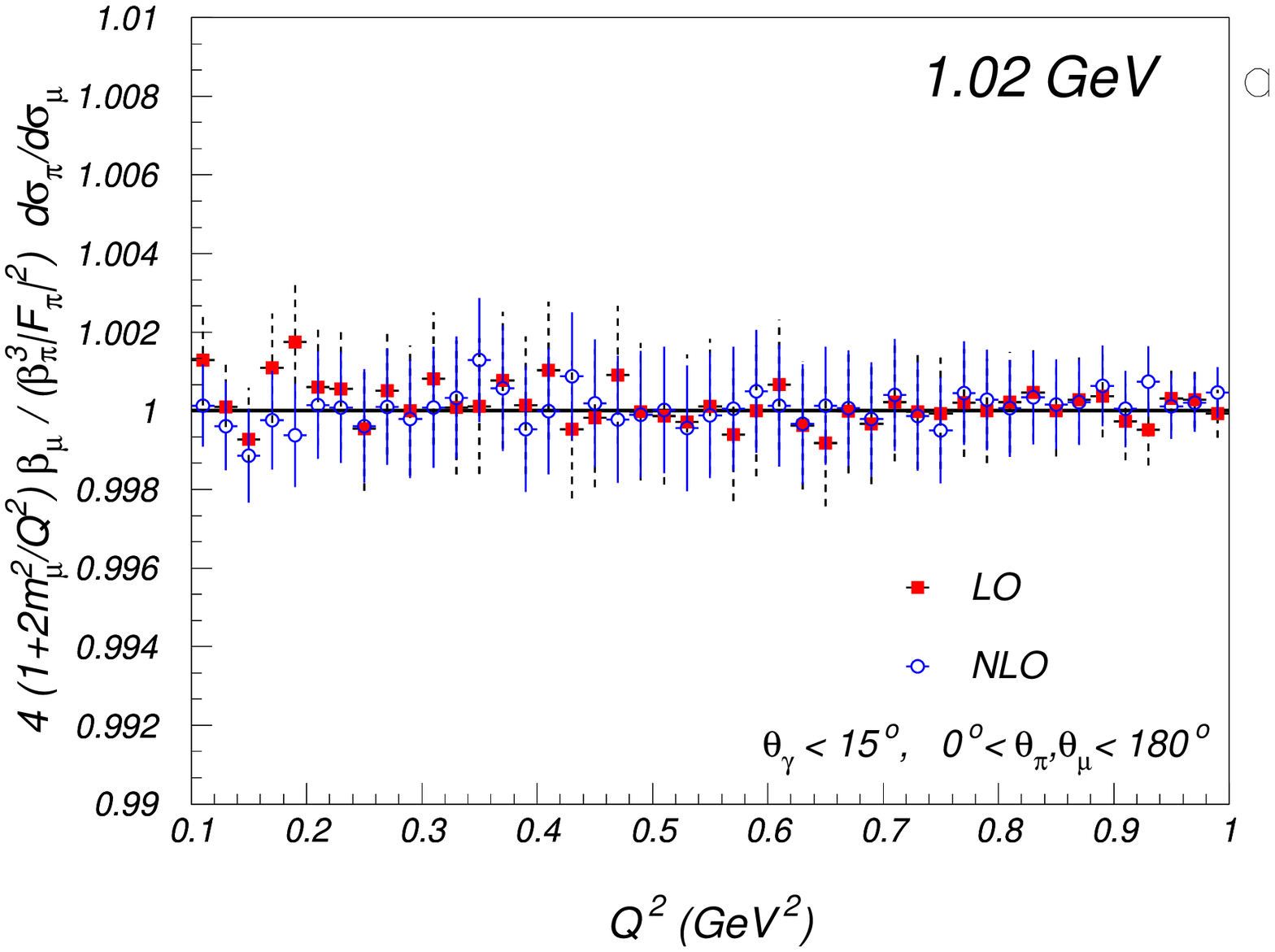,width=8.9cm}\vspace{0.3cm}
\epsfig{file=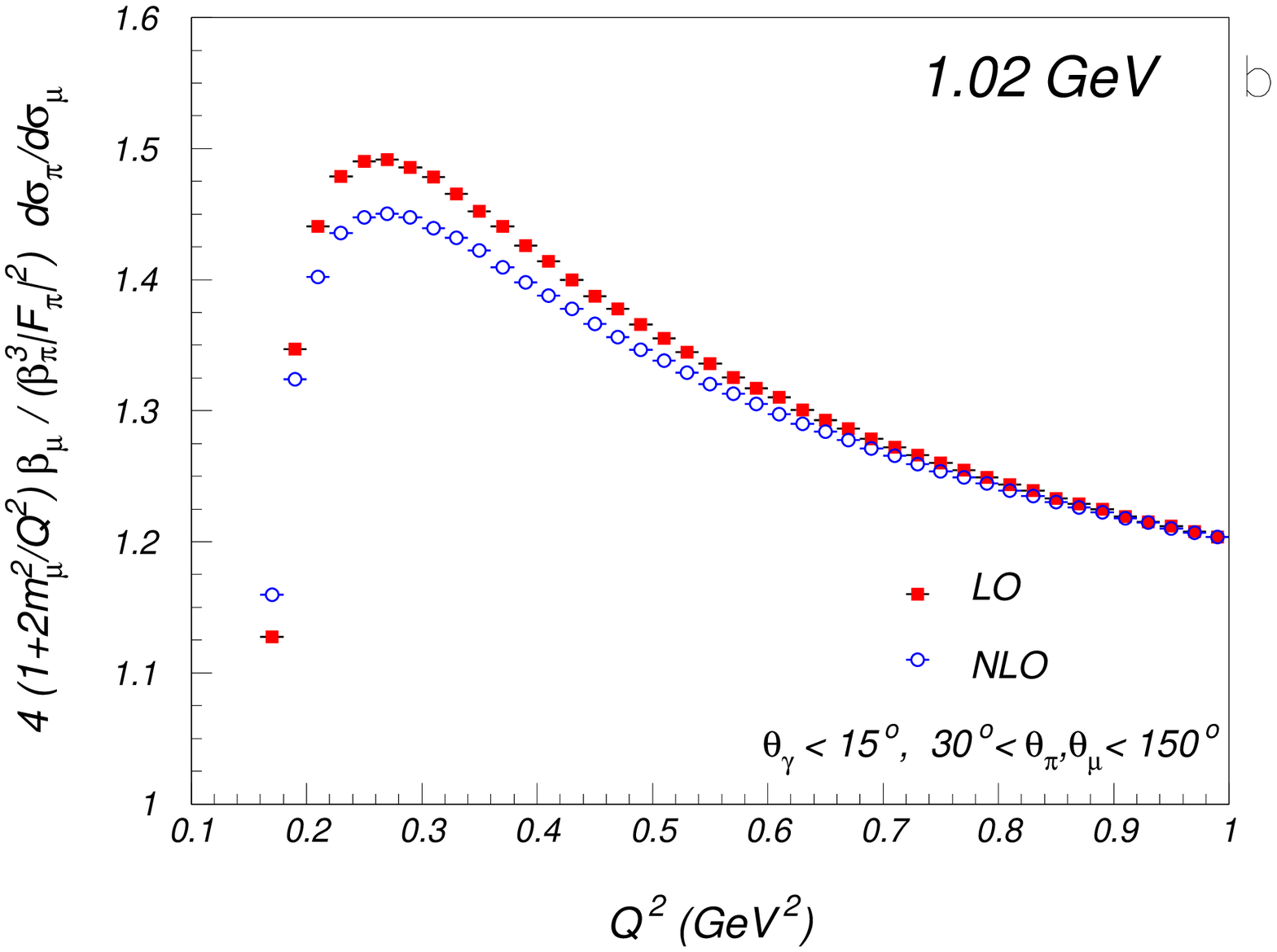,width=8.9cm}\vspace{0.3cm} 
\epsfig{file=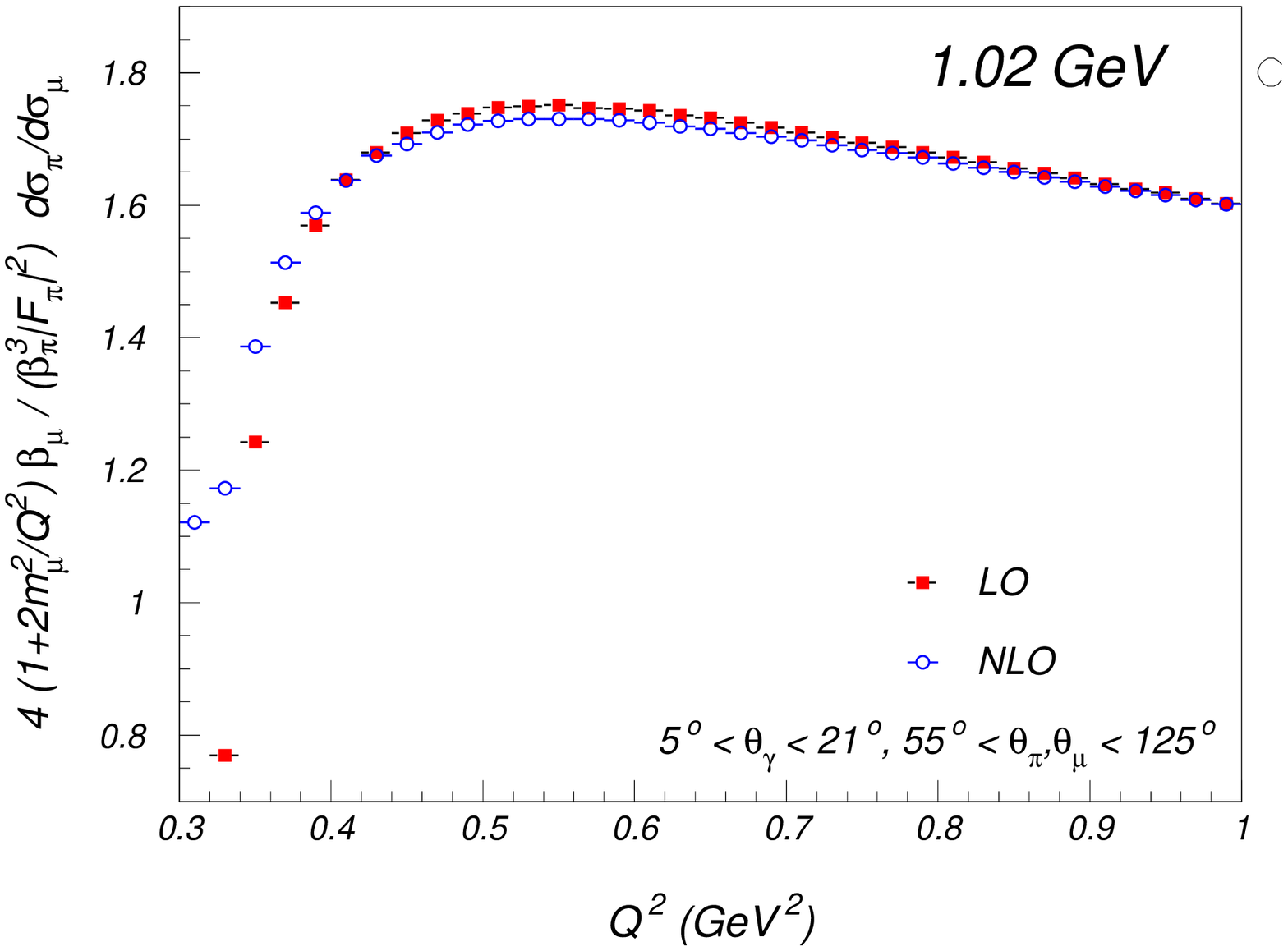,width=8.9cm} 
\end{center}
\caption{Ratio between pion and muon yields, after dividing through 
their respective R-ratio. a: no cuts on pion and muon angles;
b: with angular cuts on pion and muon angles;
c: with angular cuts on pion and muon angles and `large' angle photons.}
\label{fig:pipimumu}
\end{figure}

The R-ratio can in principle be deduced either from the measurement 
of the hadronic cross section, which requires a precise control of the
luminosity, or from the ratio between hadronic and \(\mu^{+}\mu^{-}\) 
event rates. Various radiative corrections, e.g. from the running
of the electromagnetic coupling and from ISR cancel in the ratio
between the hadronic and \(\mu^{+}\mu^{-}\) rates. Indeed, one obtains by
construction unity, if one considers the properly normalised ratio

\bea
\rho_{\pi\mu}\equiv
\frac{4(1+2m_{\mu}^{2}/Q^{2})\beta_{\mu}}
{\beta_{\pi}^{3}\mid F_{\pi}\mid^{2}} \ \
 \frac{d\sigma_{\pi^+\pi^-\gamma(\gamma)}}
 {d\sigma_{\mu^+\mu^-\gamma(\gamma)}} \ \ ,
\label{rat}
\eea

\noindent
where 

\bea
\beta_i = \sqrt{1-\frac{4m_i^2}{Q^2}} \ , \ \ i = \pi,\mu \ \ , \nonumber
\eea

\noindent
and \(F_{\pi}\) is the pion form factor.

\begin{figure*}[ht]
\vspace{-.8cm}
\begin{center}
\epsfig{file=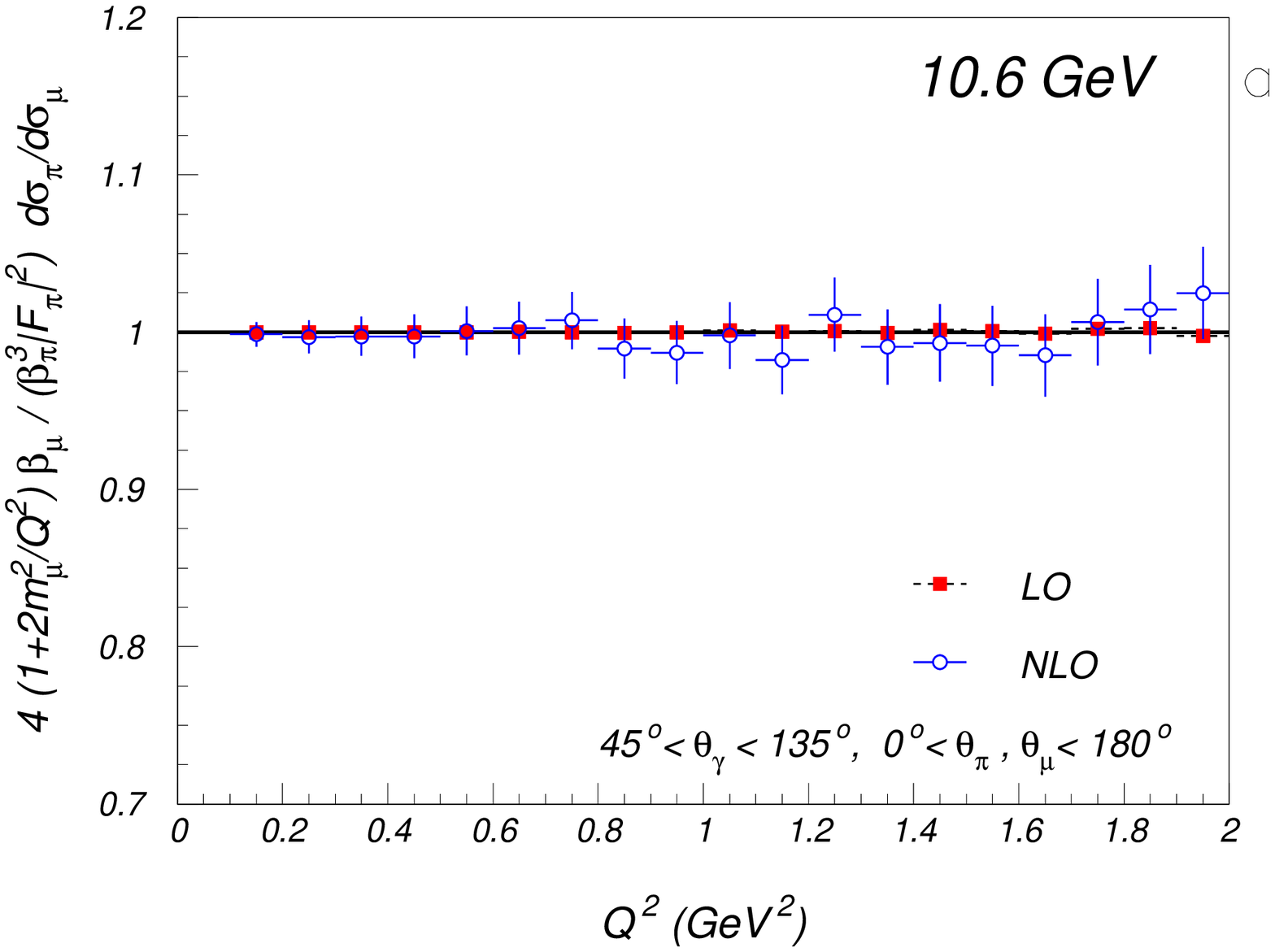,width=8.5cm}\hskip 0.6 cm 
\epsfig{file=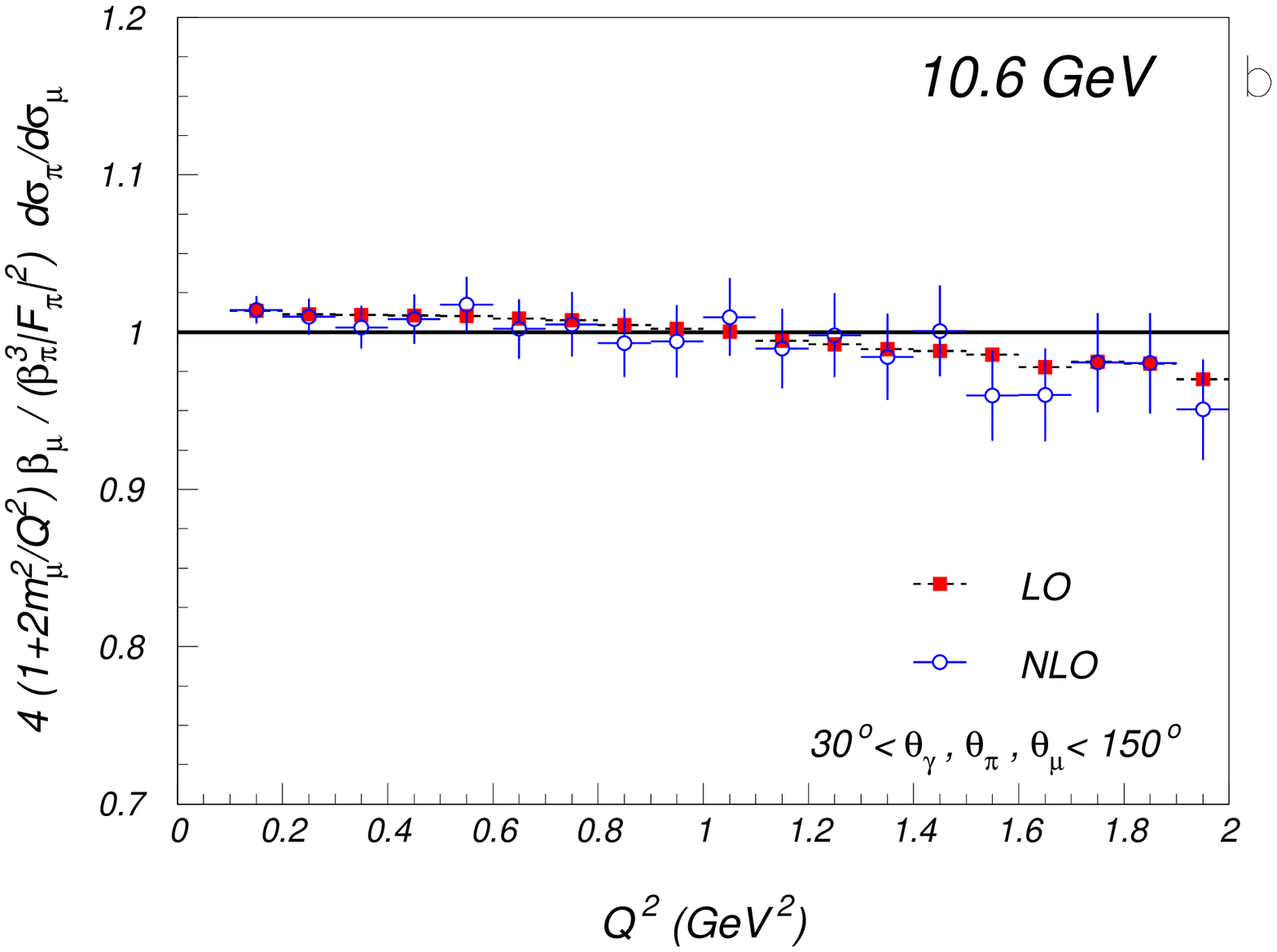,width=8.5cm} 
\end{center}
\caption{Ratio between pion and muon yields, after dividing through 
their respective R-ratio. a: no cuts on pion and muon angles;
b: with angular cuts on pion and muon angles. }
\label{fig:pipimumu1}
\end{figure*}

The result \(\rho_{\pi\mu} = 1\) 
is independent of the restrictions on the photon angular region and
is true in Born and NLO approximations.
The phase space of hadronic and \(\mu^{+}\mu^{-}\) final states,
 however, must be fully
integrated (Fig. \ref{fig:pipimumu}a). 
For realistic cuts on pion and muon angles the ratio
deviates significantly from 1,
 a consequence of their markedly different
angular distributions. The size of this effect depends on the details
 of the cuts on photon, pion and muon angles as demonstrated in 
 Figs. \ref{fig:pipimumu}b and \ref{fig:pipimumu}c.
In both figures, one observes
  a significant,
 few per cent, difference between Born and NLO predictions for 
 \(\rho_{\pi\mu}\),  depending on the details of the cuts on
 the photon and charged particle angles.
 At 10.6 GeV the ratio \(\rho_{\pi\mu}\) is of course again equal to 1
 if pions and muons are fully integrated (Fig. \ref{fig:pipimumu1}a).
 In contrast to the situation at lower energies, the inclusion of realistic
 cuts does not alter this picture drastically, a consequence of the high
 correlation between photon and pion or muon angles: photon and charged
 particles are essentially emitted back to back (Fig. \ref{fig:pipimumu1}b). 

 The shape of these curves depends only on the 
 pion and muon angular distributions,
 but not on the form factor itself. These results can thus be directly
 used to deduce efficiencies of specific experimental cuts in a 
 model-independent way, since pion and muon angular distributions are fixed
 by general considerations. For more complicated final states 
 (e.g. \(4\pi\), \(KK\pi\), \(\cdots\)) the corresponding ratio would,
 instead of \(|F_\pi|^2\beta_\pi^3/4\), directly
 involve the corresponding R-ratio, if no cuts on the hadrons are applied.
 Otherwise the results depend on the model for the hadronic form factor 
 implemented in the program.
 An important advantage of the radiative
return is implicit in all these considerations: by measuring \(Q^{2}\) 
 directly,
 the invariant squared mass of the hadronic final state,
one has direct access to R at the corresponding value of \(Q^{2}\).
This differs from the measurement of the inclusive cross section 
as a function of \(\sqrt{s}\) (energy scan). To extract the true R(\(s\)), an 
unfolding has to be performed, which requires in principle 
 the knowledge of the cross section over the full energy range below
and a precise knowledge of the radiator function. 
 In contrast, using the radiative return method, it is still necessary
 to know the QED radiator function, 
 but no unfolding is required as one measures the \(Q^2\) of the
 hadronic system, and thus has `access' to the hadronic cross
 section at that given \(Q^2\).

\begin{figure}[htb]
\epsfig{file=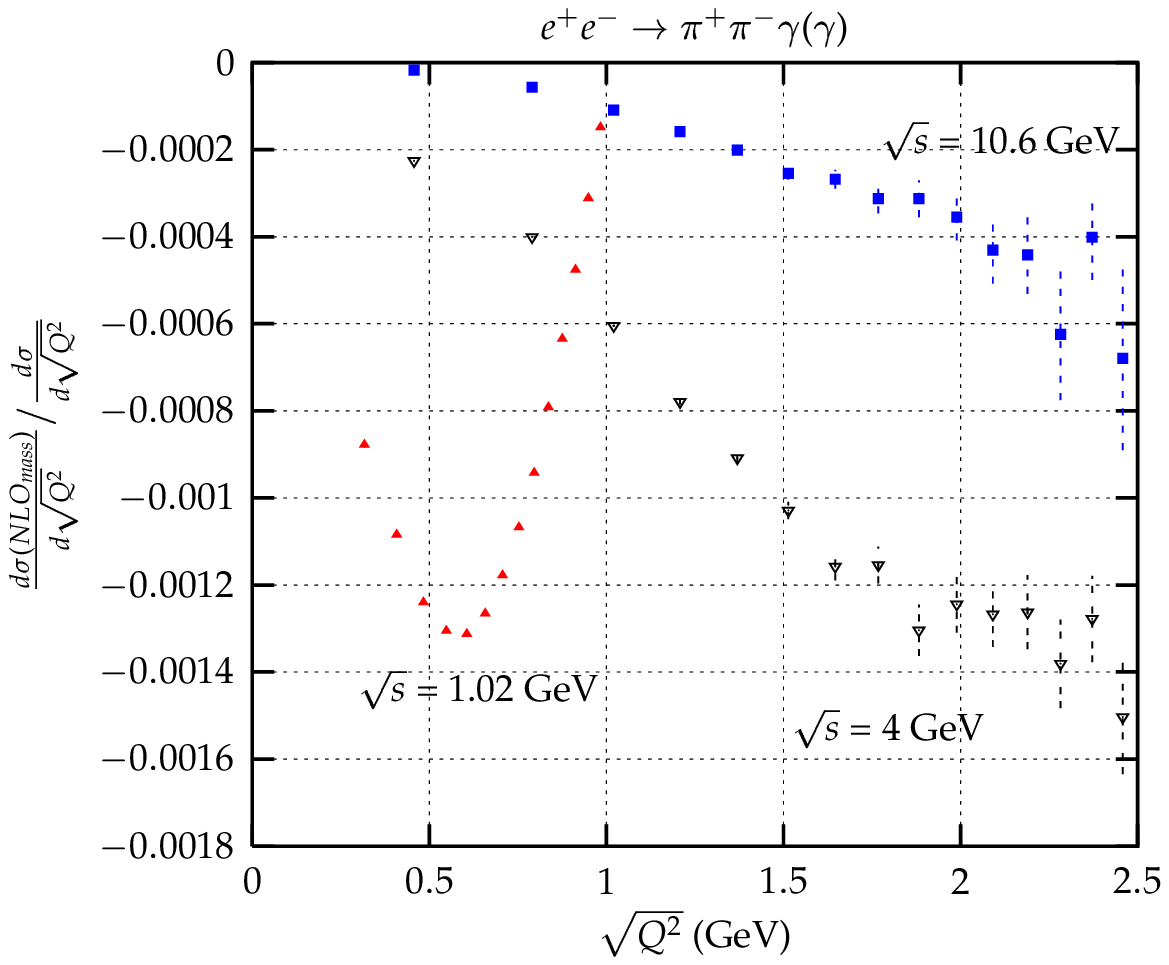,width= 8.5 cm,}
\caption{The relative contributions of the non-leading mass corrections to the
 differential cross section at \(\sqrt{s}\) = 1, 4 and 10 GeV. }
\label{fig:nlo_mass}
\end{figure}

Let us discuss those \(\alpha m_{e}^{2}/s\) and 
\(\alpha (m_{e}^{2}/s)^{2}\) terms, which in combination with their 
singular angular dependence integrate to corrections of order
\(\alpha/\pi\). These are included in the present version of
 PHOKHARA. The corresponding leading order corrections 
 proportional to \(m_e^2\) are typically of the order of a few per cent
 \cite{Rodrigo:2001cc}, while the non-leading ones are of order 0.1\%.
 This can be
 seen from Fig. \ref{fig:nlo_mass}. The size of these effects
 is consistent with the expectations
 for \(\alpha/\pi\) terms without logarithmic enhancement.
 These terms will become important when the  precision of
 the measurement will be below 1\%. 
Their proper treatment is in that
 case crucial as they do depend on \(Q^2\) and do change 
 the \(Q^2\) distribution from which the hadronic
 cross section is extracted.                             

\section{Initial versus final state radiation}

\begin{figure}[ht]
\begin{center}
\vspace{-.8cm}
\epsfig{file=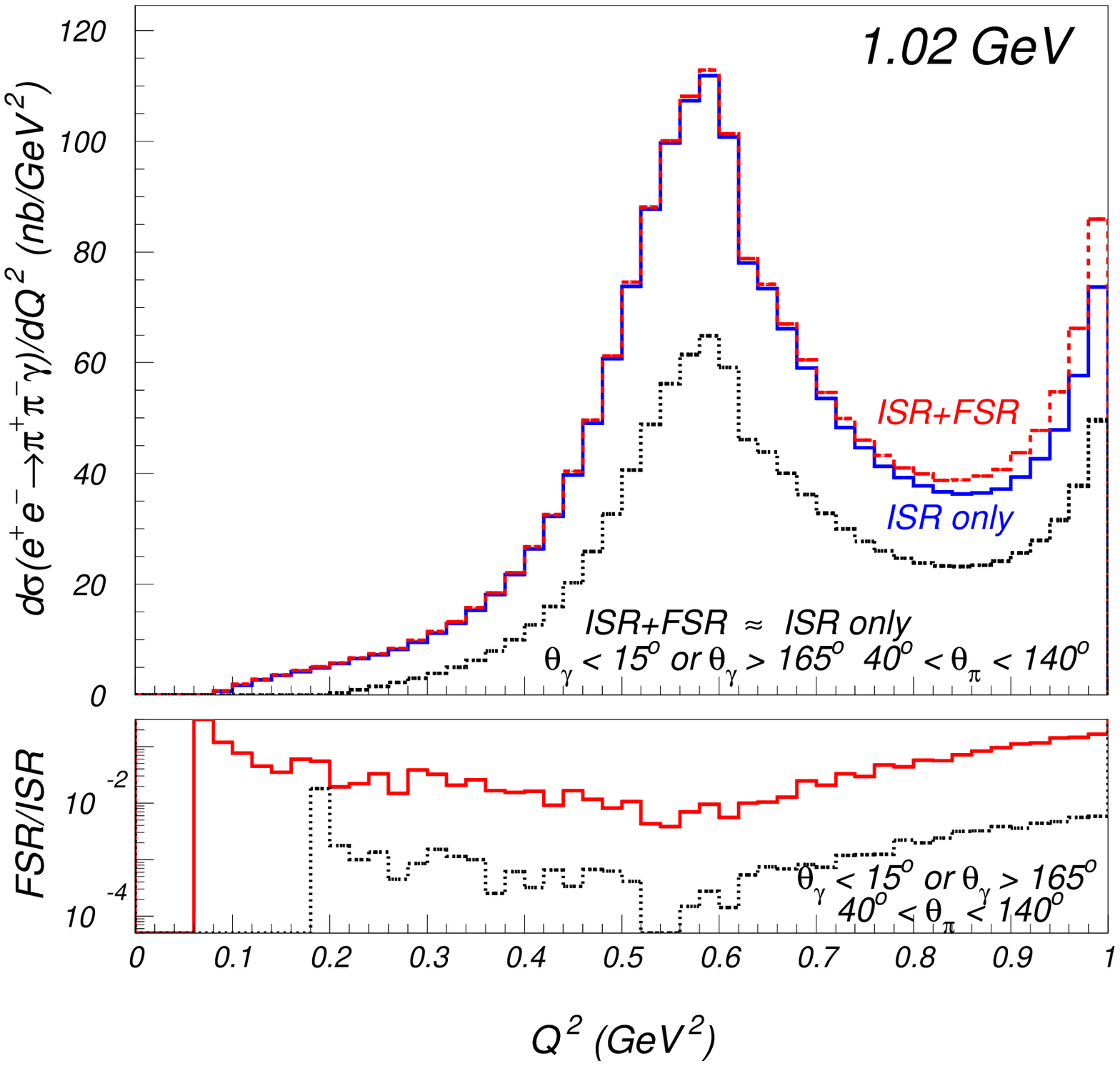,width=9 cm,}\vspace{-3. cm}
\caption{ The suppression of the FSR contributions
to the cross section by a suitable choice of angular cuts.
 Results from the PHOKHARA generator.
No cuts (upper curves) and suitable cuts applied (lower curves). }
\label{fig:isrtofsr}
\end{center}
\end{figure}                                                                   

 A potential complication for the
measurement of the pion form factor or generally of the R-ratio
may arise from the interplay between photons from ISR and 
FSR. Their relative strength is strongly dependent
on the photon angle relative to the beam and the pion directions,
the cms energy of the reaction
and the invariant mass of the hadronic system. 
 FSR from hadronic final states cannot be predicted from first principles
and thus has to be modelled. 
The model amplitude can nevertheless be tested by considering 
charge-asymmetric differential distributions, which arise from the interference
between ISR and FSR amplitudes \cite{Binner:1999bt}.
 In leading order the complete matrix
element squared is given by 

\begin{equation}
|{\cal M}|^2 = |{\cal M}_{\mathrm{ISR}}|^2  + |{\cal M}_{\mathrm{FSR}}|^2
+2 \mathrm{Re}[{\cal M}_{\mathrm{ISR}} {\cal M}_{\mathrm{FSR}}^\dagger]~,
\label{eq:interference}
\end{equation}
which is still independent of the model for FSR.

FSR and its interference with ISR were already included in 
EVA~\cite{Binner:1999bt} for the two-pion case. 
The pions were assumed to be point-like, and scalar QED was 
applied to simulate photon emission off the charged pions. 
It was demonstrated there that ISR dominates for suitably chosen 
final states, namely 
those with hard photons at small angles relative to the beam, well 
separated from the pions. FSR can therefore 
be reduced to a reasonable limit, and moreover, can be controlled 
by the simulation (see also \cite{CDKMV2000}).
Similar results can be obtained 
using the new version of PHOKHARA, were FSR and ISR--FSR interference
 are included 
for two pions and muons (see Appendix A for details)
at LO. The photon emission from pions is again modelled by a point--like
 pion-photon interaction.
The upper two curves of Fig. \ref{fig:isrtofsr} describe the differential
 cross section with  arbitrary photon and pion angles. The contribution
 of FSR is clearly visible. Once photon emission is restricted to angles
 close to the beam and if the pion- and photon-allowed angular ranges 
 do not overlap (lower curve), FSR is clearly negligible.

\begin{figure}
\begin{center}
\vspace{-.8cm}
\epsfig{file=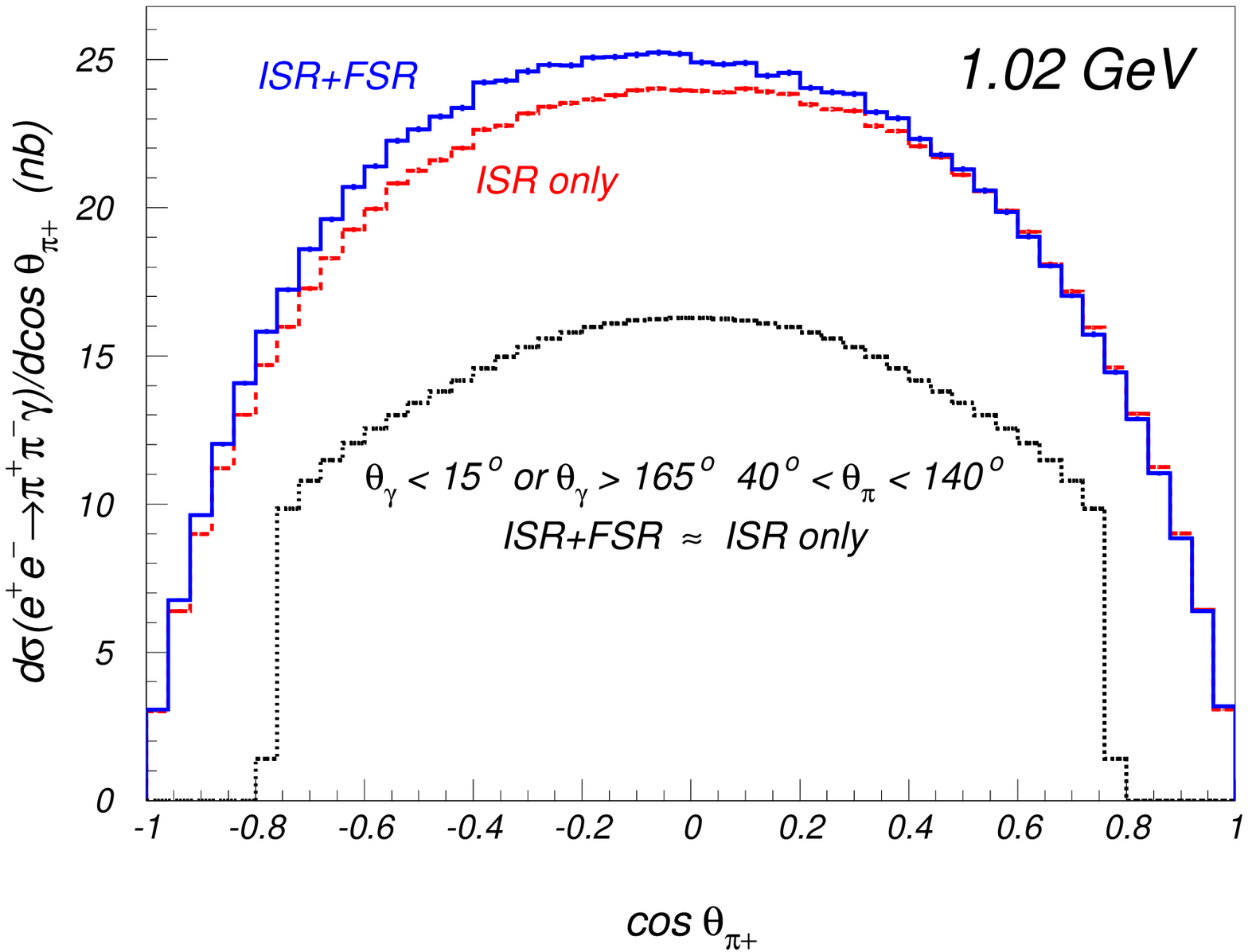,width=9cm} 
\end{center}
\caption{Angular distribution of $\pi^+$ with and without FSR 
for different angular cuts.}
\label{fig:angular}
\end{figure}

\begin{figure}
\begin{center}
\vspace{-.8cm}
\epsfig{file=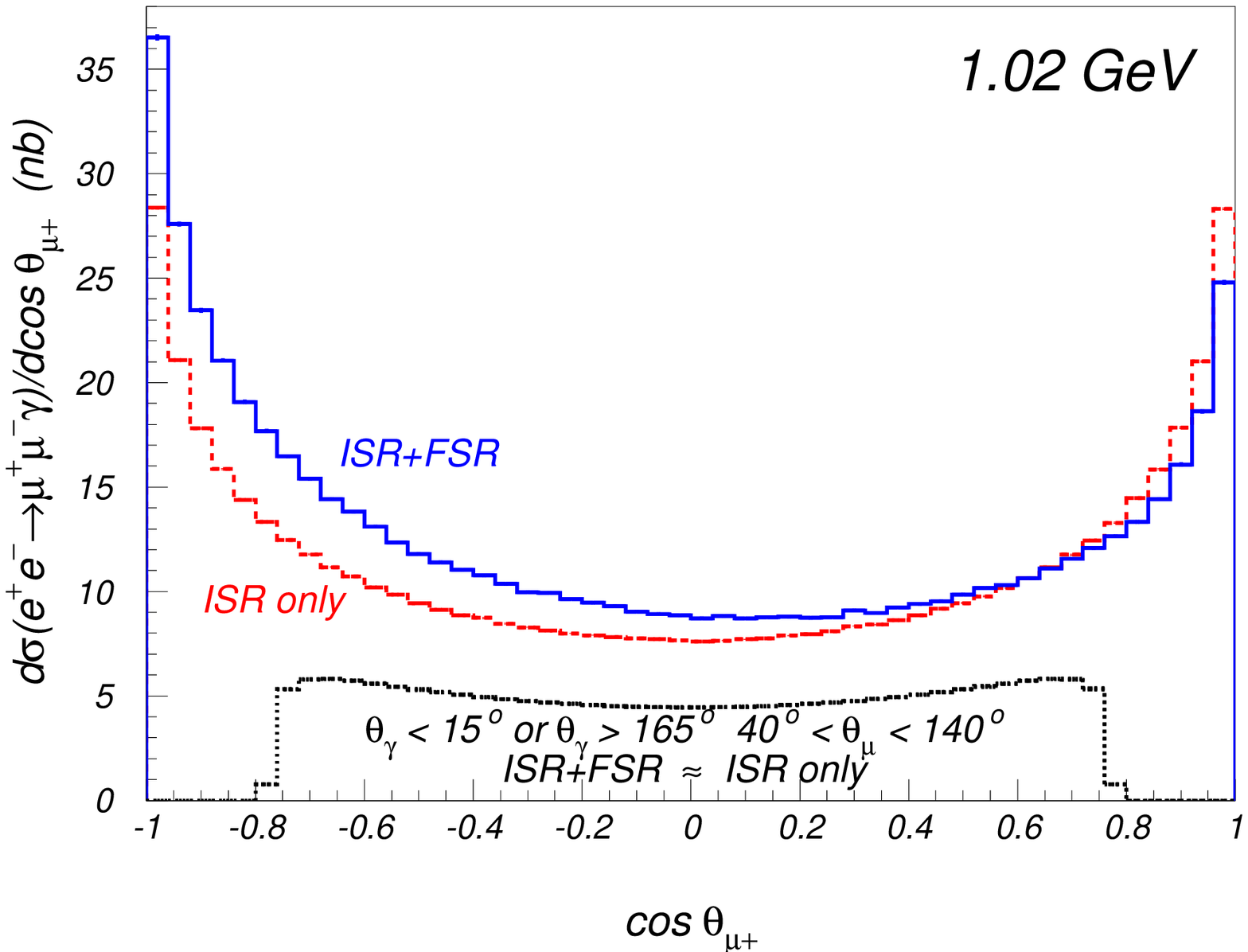,width=9cm} 
\end{center}
\caption{Angular distribution of $\mu^+$ with and without FSR 
for different angular cuts.}
\label{fig:mangular}
\end{figure}

The third term in the right-hand side of Eq.~(\ref{eq:interference}),
ISR--FSR interference, is odd under charge conjugation, and its
contribution vanishes after angular integration.
It gives rise, however, to a relatively large charge asymmetry and, 
correspondingly, to a forward--backward asymmetry
\begin{equation}
A(\theta) = \frac{N^{\pi^+}(\theta)-N^{\pi^+}(\pi-\theta)}
{N^{\pi^+}(\theta)+N^{\pi^+}(\pi-\theta)}~.
\end{equation}
The asymmetry can be used for calibration of the FSR amplitude,
and fits to the angular distribution \(A(\theta)\) can test
details of its model dependence. Given sufficiently large event rates
this procedure can be performed for different \(\theta_\gamma\), thus
allowing for an unambiguous reconstruction of the FSR amplitude.

This is illustrated in Figs.~\ref{fig:angular} and~\ref{fig:mangular},
where the angular distributions of $\pi^+$ and $\mu^+$ respectively are 
shown for different kinematical cuts. The angles are defined with respect 
to the incoming positron. If no angular cut is applied, 
the angular distribution in both cases is highly asymmetric as 
a consequence of the ISR--FSR interference contribution. If cuts 
suitable to suppress FSR, and therefore the ISR--FSR interference,
are applied, the distributions become symmetric. 

 These investigations can also be performed for different photon energies,
 thus exploring FSR in different regions of \(Q^2\). We will return
 to this aspect in a future publication.

\begin{figure}
\begin{center}
\vspace{-.8cm}
\epsfig{file=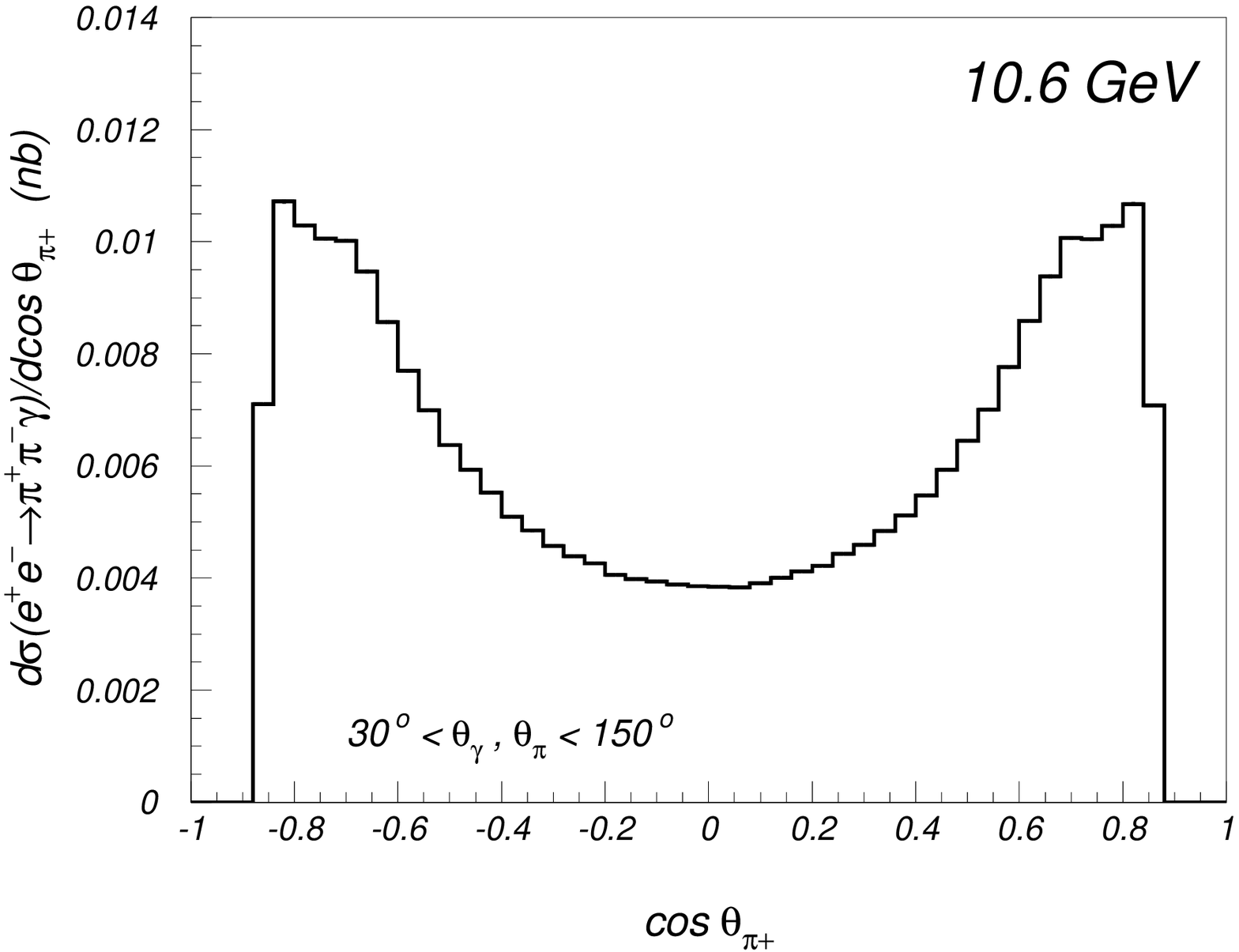,width=9cm} 
\end{center}
\caption{Angular distribution of $\pi^+$ at $\sqrt{s}$=10.6~GeV 
(ISR \(\simeq\) FSR+ISR).}
\label{fig:angular10GeV}
\end{figure}

At B-factories, where one has to deal with very hard tagged photons,
the kinematic separation between the 
photon and the hadrons becomes very clear.
For events where hadrons and photon are produced mainly back to back, 
the suppression of FSR is naturally accomplished and no special 
angular cuts are therefore needed to control FSR versus ISR 
at higher energies (Fig. \ref{fig:angular10GeV}). The relative size
of the FSR is of the order of a few per mil, but does depend on the 
value of the pion form factor at \( \sqrt{s}\) = 10 GeV, which is
 extrapolated from the low energy data. 

\begin{figure}
\begin{center}
\vspace{-.8cm}
\epsfig{file=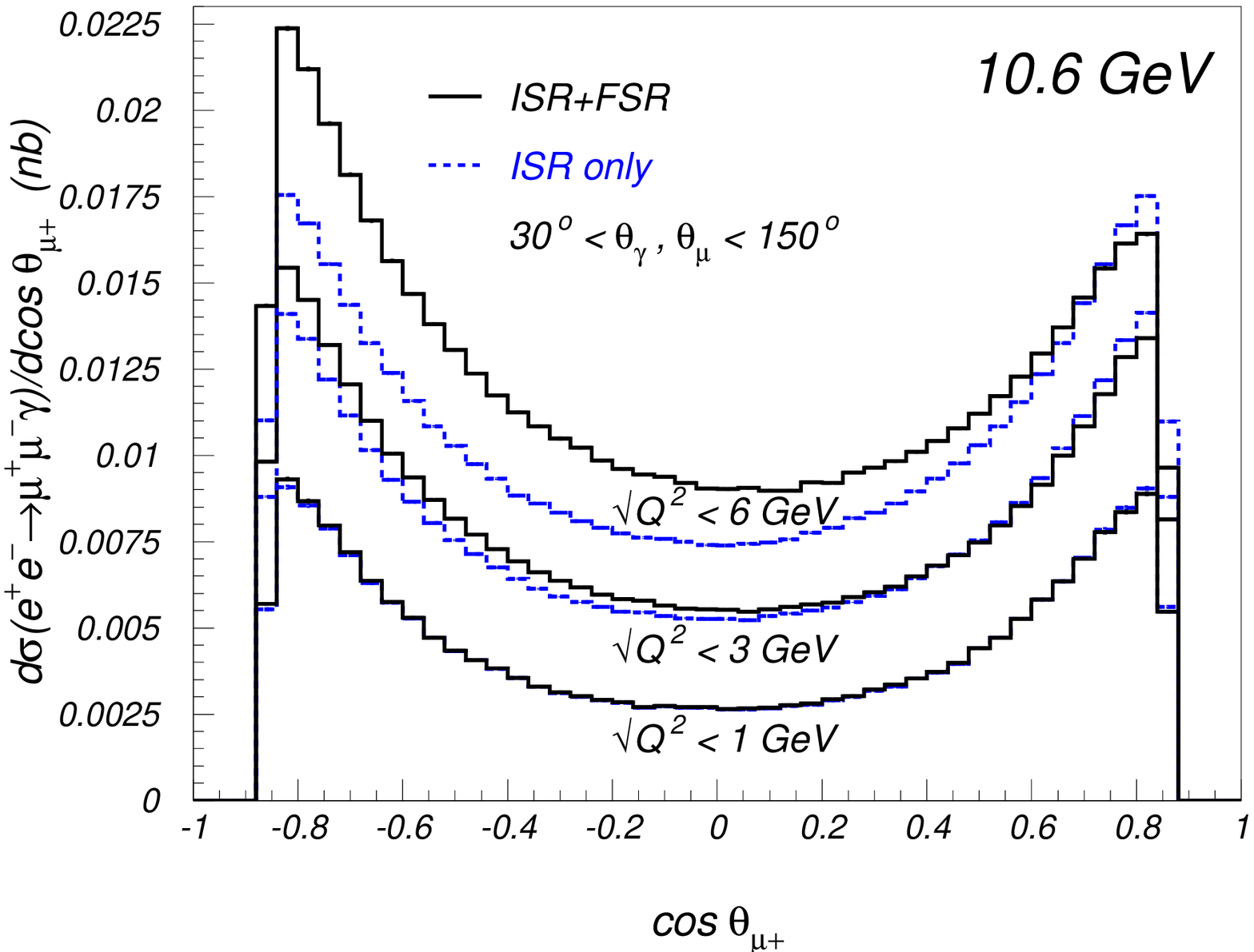,width=9cm} 
\end{center}
\caption{Angular distribution of $\mu^+$ at $\sqrt{s}$=10.6~GeV
 for various \(Q^2\) cuts.}
\label{fig:mangular10GeV}
\end{figure}

 The suppression of FSR contributions to \(\pi^+\pi^-\gamma\) events
 is also a consequence of the rapid decrease of the form factor above
 \(\sim\) 1 GeV. It is therefore instructive to study the corresponding
 distributions for \(\mu^+\mu^-\gamma\) final states. For 
 \(\sqrt{Q^2} \le\) 1 GeV FSR is still tiny.
 Around 3 GeV a small charge-asymmetric
 interference term becomes visible (Fig. \ref{fig:mangular10GeV}),
 which is still irrelevant after
 averaging over \(\mu^+\) and  \(\mu^-\). At large \(Q^2\), however,
 FSR plays an important role both for the charge-asymmetry and the
 charge symmetric term.

\section{The four-pion mode}

\begin{figure}[htb]
\begin{center}
\vspace{1cm}
\epsfig{file=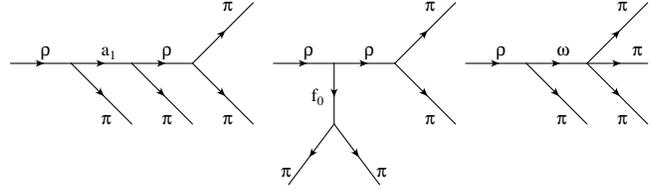,width=8.5cm} 
\end{center}
\caption{Diagrams contributing to the 4$\pi$ hadronic current.}
\label{fig:4picurrent}
\end{figure}

Because of the modular structure of PHOKHARA, additional hadronic 
modes can be easily implemented. The four-pion channels
($2\pi^+ 2\pi^-$ and $2\pi^0\pi^+\pi^-$), which give the dominant 
contribution to the hadronic cross section in the region from 1 to 2~GeV,
are a new feature of our event generator.

\begin{figure}[htb]
\epsfig{file=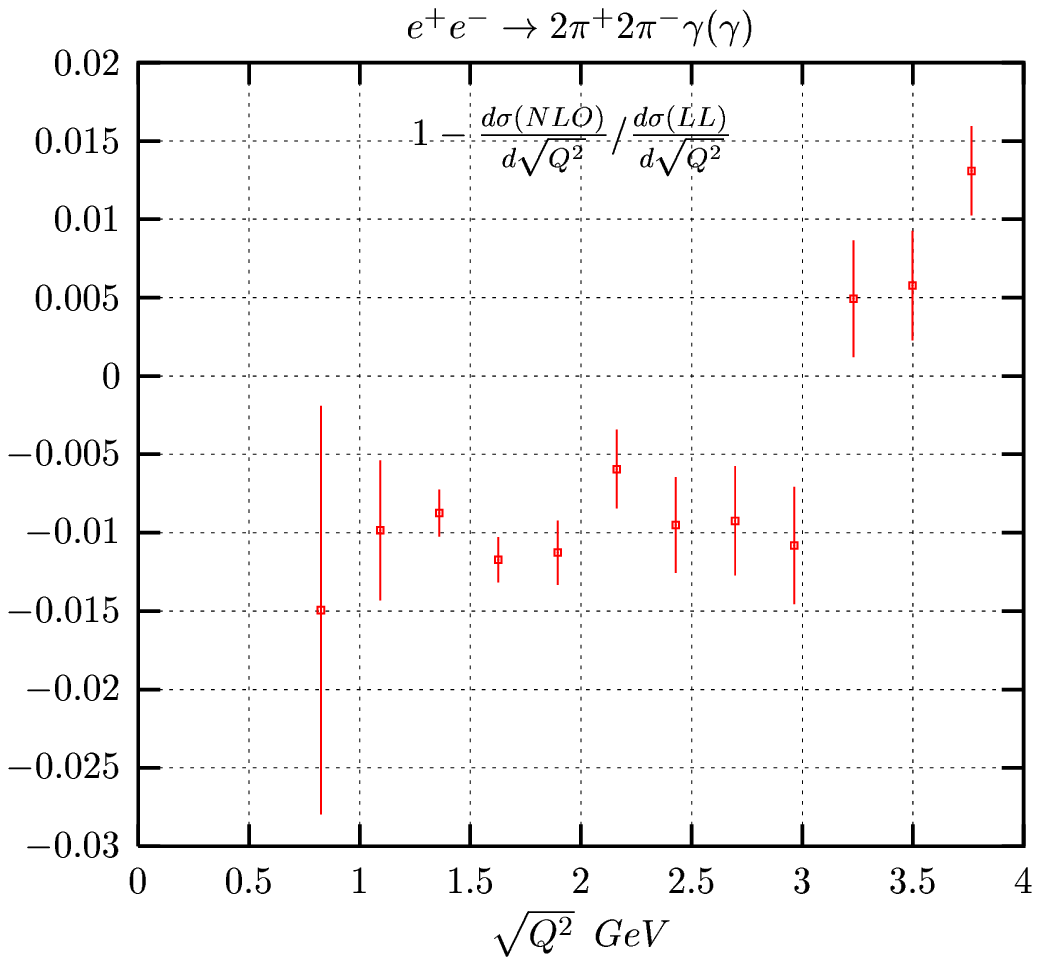,width=8.5 cm,}
\caption{The relative
non-leading contributions to the differential cross section
at \(\sqrt{s}\) = 4 GeV. NLO - full next-to-leading result,
LL - leading logarithmic approximation.  }
\label{fig:ll_nlo_1}
\end{figure}
\begin{figure}[htb]
\epsfig{file=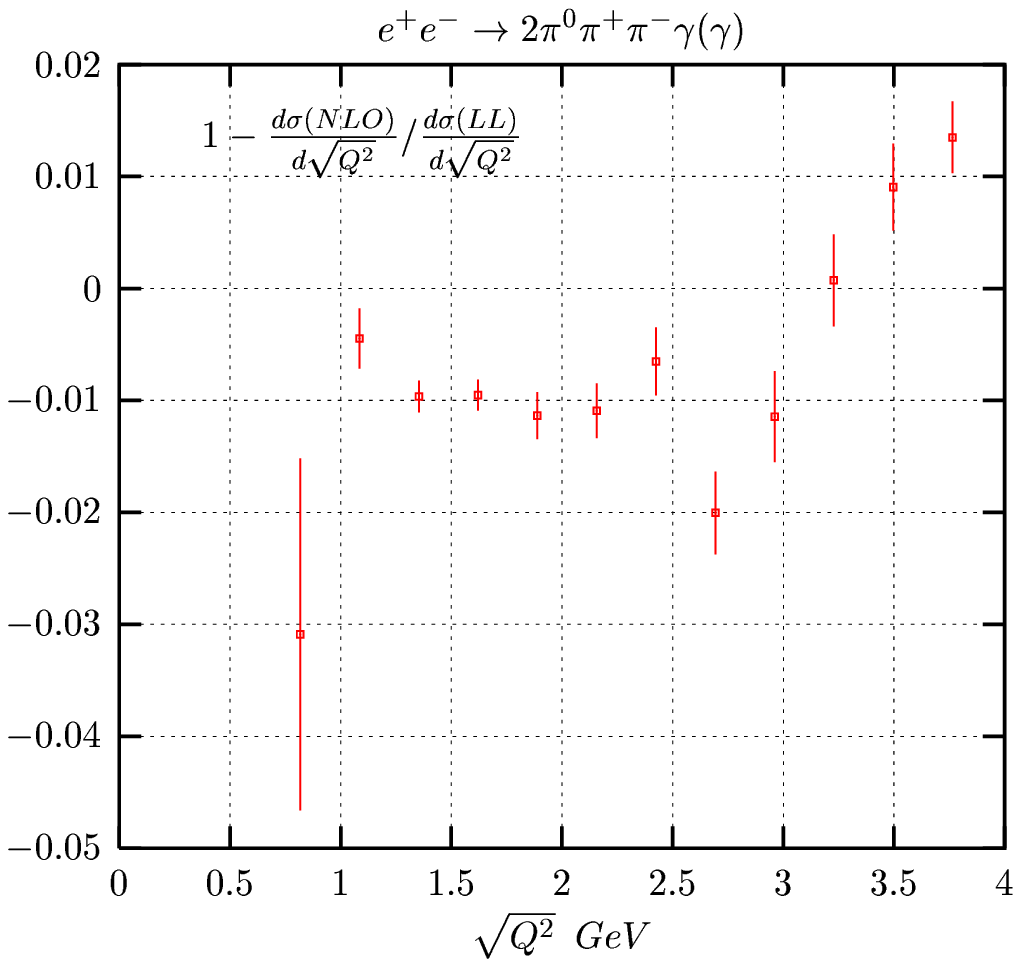,width=8.5 cm,}
\caption{The relative
non-leading contributions to the differential cross section
at \(\sqrt{s}\) = 4 GeV. NLO - full next-to-leading result,
LL - leading logarithmic approximation.}
\label{fig:ll_nlo_2}
\end{figure}                                                                   

 Isospin invariance relates the amplitudes of the 
$e^+e^- \to 2\pi^+ 2\pi^-$ and $e^+e^- \to 2\pi^0\pi^+\pi^-$ processes 
and those for $\tau$ decays into $\pi^-3\pi^0$ and $\pi^+2\pi^-\pi^0$
\cite{K2,Czyz:2000wh}.
The description of the four-pion hadronic current 
follows~\cite{Czyz:2000wh,Decker:1994af}. The basic building blocks
of this current are schematically depicted in Fig.~\ref{fig:4picurrent}
and described in detail in~\cite{Czyz:2000wh}.

Results obtained with PHOKHARA for these channels have been 
compared with the Monte Carlo, which simulates the same 
process at LO~\cite{Czyz:2000wh} and includes additional 
collinear radiation through the SF technique. Typically, differences 
of order 1$\%$ are found (see Figs.~\ref{fig:ll_nlo_1} and
 ~\ref{fig:ll_nlo_2}), which are of the 
expected size and of the same order as for the two-pion final 
state~\cite{RCKS}.

The generation
 of the pion four momenta is however different from the one described
 in \cite{Czyz:2000wh}. In the present version
 of the program we absorb the most prominent peaks in the
 four-pion hadronic current to obtain a more efficient Monte Carlo
 generation. The \(Q^2\) distribution is peaked 
 around $\sqrt{Q^2}$ = 1.5 GeV,
 with a large width of \(\sim\) 0.5 GeV. This is the result of
 an interplay between several resonances present in that region.
 Nevertheless one 
 Breit--Wigner resonance provides an adequate approximation for 
 efficient generation. For the approximant and the generation
 of the \(Q^2\) distribution we use
\begin{align}
 &f_3(Q^2) = \frac{s}{s-Q^2}
 +\frac{s^2}{(Q^2-m^2)^2
 + \Gamma^2 m^2} \ ,
\label{a3}
 \end{align}

\noindent
with \(m=1.5\) GeV and \(\Gamma=\) 0.5 GeV.

It takes care of soft photon emission (\(s - Q^2 \sim E_\gamma\) )
 and the aforementioned resonant 
behaviour. For the process \(e^+e^- \to 2\pi^0\pi^+\pi^- \gamma (\gamma)\)
we furthermore absorb 
the \(\omega\) peaks in the four momentum squared 
 \(Q_{0+-}^2=p_\omega^2= (p_0+p_+ + p_-)^2 \).
 The
approximant used for that purpose, according to which the three-particle
four momenta squared are generated, reads

\begin{align}
 &f_4(Q_{134}^2,Q_{234}^2) = 2 
 \non \\ &
+ \frac{m_{\omega}\Gamma_{\omega}}{(Q_{134}^2-m_{\omega}^2)^2
 + \Gamma_{\omega}^2 m_{\omega}^2}
+ \frac{m_{\omega}\Gamma_{\omega}}{(Q_{234}^2-m_{\omega}^2)^2
 + \Gamma_{\omega}^2 m_{\omega}^2} \ ,
\label{a4}
 \end{align}

\noindent
where \(Q_{134}\) and \(Q_{234}\) are the four momenta of the two
 \(\pi^0\pi^+\pi^-\) subsystems. The other variables are generated as 
 described in \cite{Czyz:2000wh}.

\begin{table}
\caption{Total cross section (nb) for the process
$e^+ e^- \rightarrow 4\pi \gamma$ at NLO for different values
of the soft photon cutoff at \(\sqrt{s}=  \) 1.02 GeV.
 Only initial state radiation.
 One of the photons with energy $>$ 10 MeV. $Q^2<1$GeV.
 No further cuts applied.}
\label{tab:epstest_4pi}
\begin{center}
\begin{tabular}{ccc}
$w$ & $2\pi^+2\pi^-$ &  $2\pi^0 \pi^+\pi^- $\\ \hline
$10^{-3}$ & 0.170167(15)  & 0.55725(5)\\
$10^{-4}$ & 0.170413(14)  & 0.55844(5)\\
$10^{-5}$ & 0.170431(15)  & 0.55845(5)\\ \hline
\end{tabular}
\end{center}
\end{table}                          

\begin{figure*}[!ht]
\begin{center}
\epsfig{file=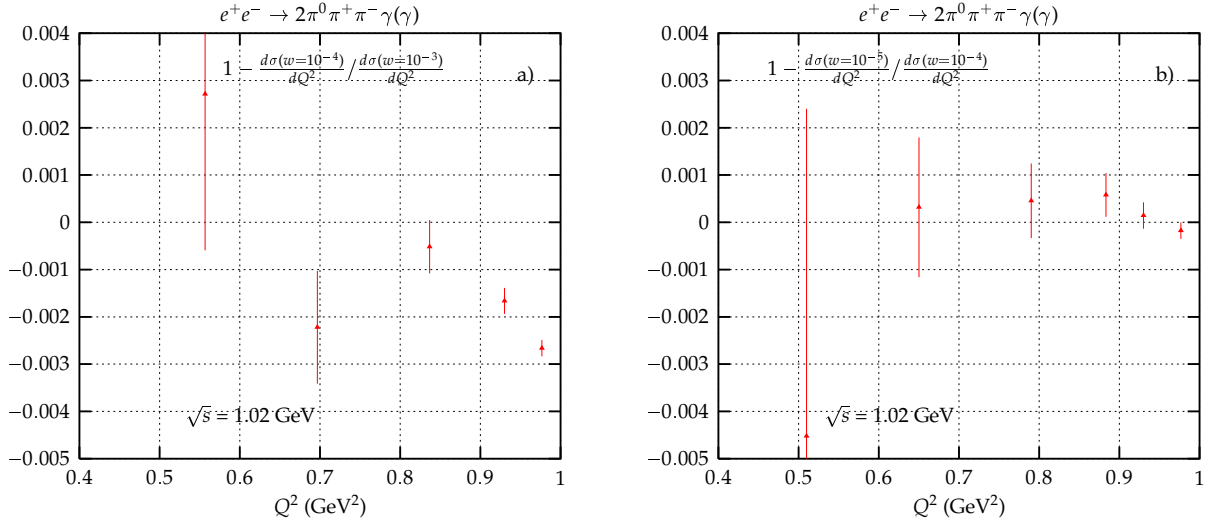,width=7.5 cm,}\hskip 1 cm
\epsfig{file=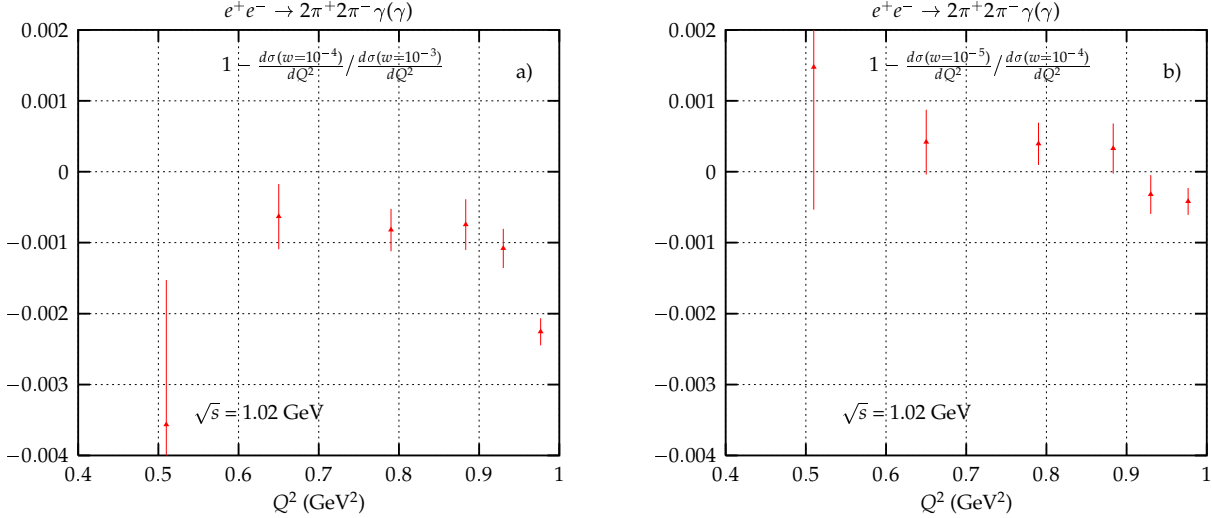,width=7.5 cm,}
\caption{The relative difference of the differential cross sections for two
 different values of the soft photon cutoff at \(\sqrt{s}\) = 1.02 GeV.
 One of the photons was required to have energy $>$ 10 MeV. 
 No further cuts were applied.}
\label{fig:e34}
\end{center}
\end{figure*}
\begin{figure*}[!ht]
\begin{center}
\epsfig{file=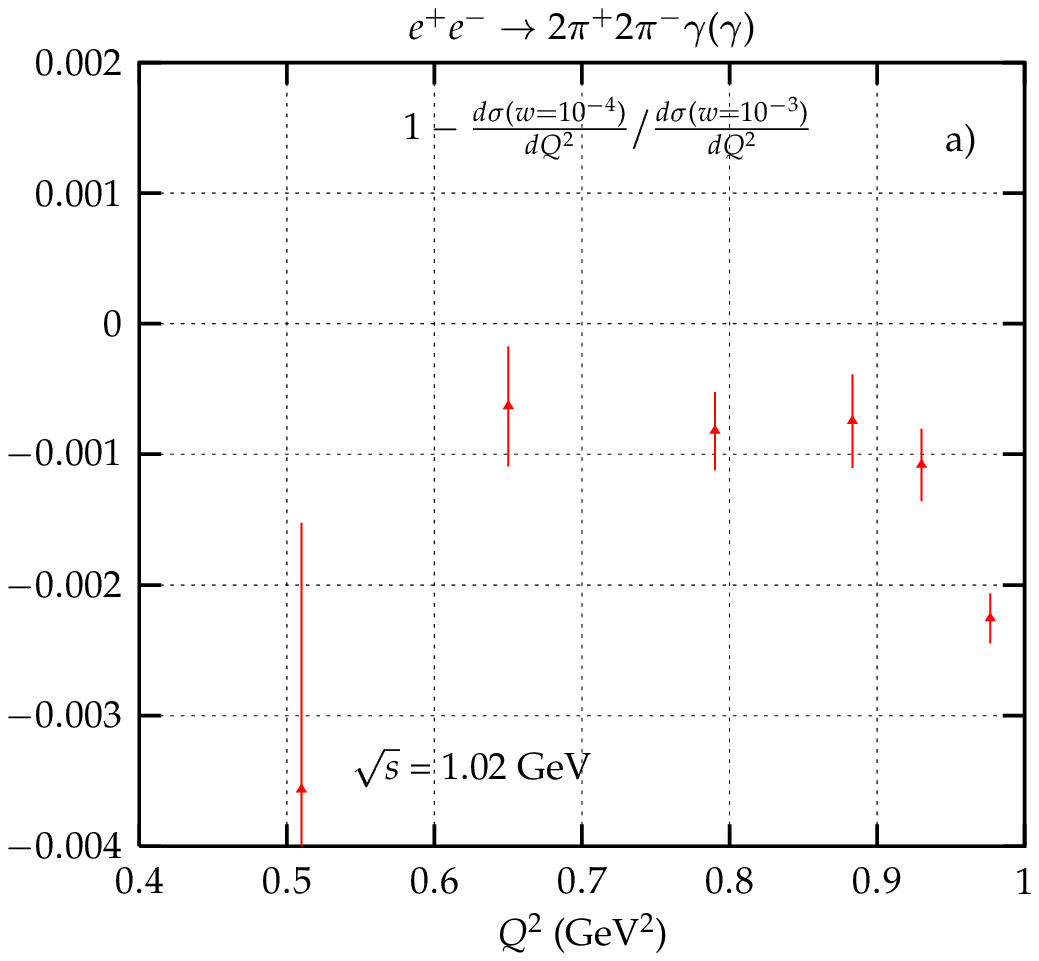,width=7.5 cm,} \hskip 1 cm
\epsfig{file=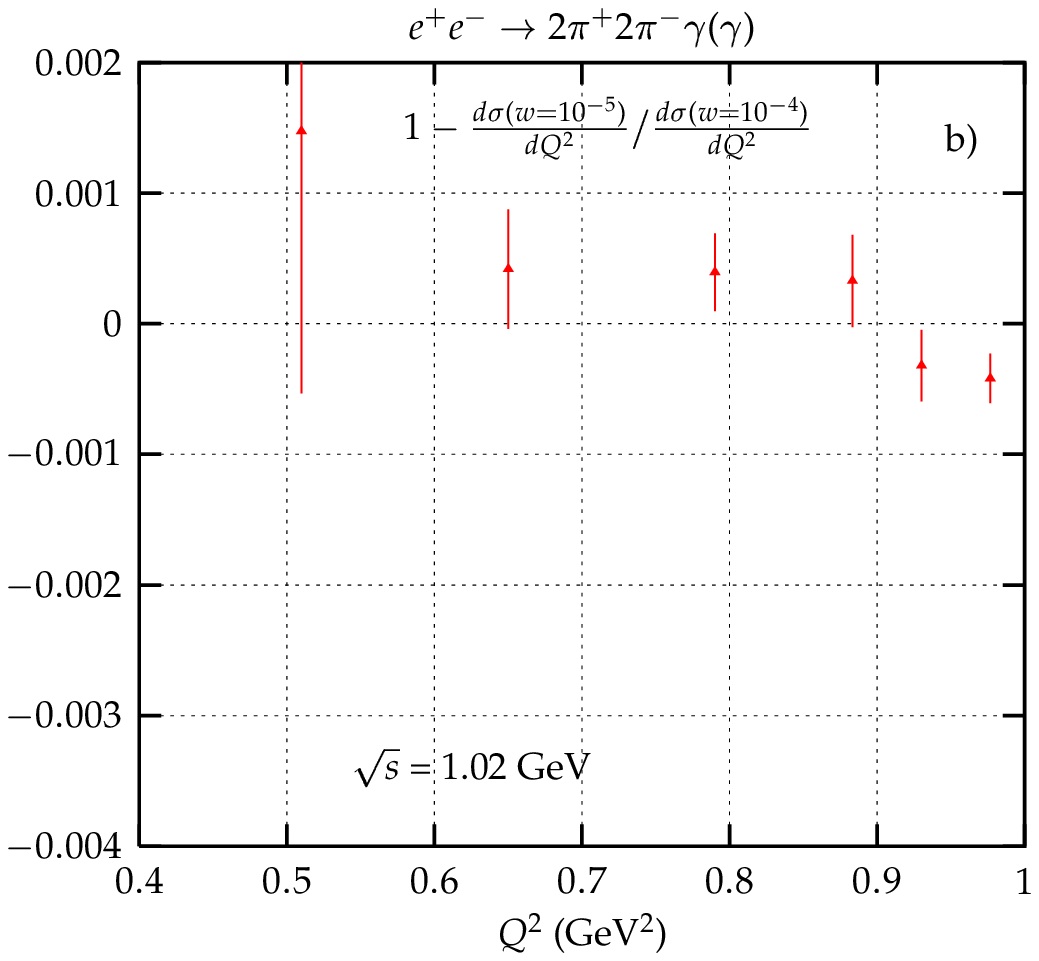,width=7.5 cm,}
\caption{The relative difference of the differential cross sections for two
 different values of the soft photon cutoff at \(\sqrt{s}\) = 1.02 GeV.
 One of the photons was required to have energy $>$ 10 MeV. 
 No further cuts were applied.}
\label{fig:e34c}
\end{center}
\end{figure*}

 Also for the four-pion modes, we have tested the independence
 of the result of the soft--hard separation parameter \(w\).
 The results are very much similar to the \(\pi^+\pi^-\) mode
 and are summarised in Table \ref{tab:epstest_4pi} and Figs. 
 \ref{fig:e34} and \ref{fig:e34c}.
  Again the choice \(w=10^{-4}\) is the proper one.

\section{Conclusions}

The Monte Carlo generator PHOKHARA, which simulates the radiative return
at electron--positron colliders, has been extended from large angles
into the collinear region using recent results for the virtual
corrections to photon emission, which are valid for all photon angles.
Comparing the program with analytical results, a technical precision
better than $ 0.5\times10^{-3} $ is demonstrated. The importance of NLO
corrections for the extraction of a correct value for the R-ratio
is emphasised. 

A number of corrections vanish if 
the ratio between hadron and muon pair cross section is considered.
 For low energies, around
1 GeV, the ratio depends strongly on the cuts on the charged particles
and corrections have to be applied. At higher energies, 
around 10 GeV, and for low $Q^2$, the dependence on these cuts is
drastically reduced.

In the new version of PHOKHARA, described in this article,
final state radiation in leading order treatment is included.
 We discuss the implications for
the measurement of the pion form factor. Suitable cuts allow, on the one hand, 
the determination of this model-dependent amplitude, on the other hand it is
possible to select configurations that are entirely dominated by initial 
state radiation. 

Finally we extend the program final states with four-pions configuration
as a first step towards the inclusion of a multitude of exclusive states.

\section*{Acknowledgements}

We would like to thank: Nicolas Berger, Stanley Brodsky,  Oliver 
Buchm\"uller and Dong Su for very interesting discussions, the 
members of KLOE collaboration for their continued interest in the subject,
and Achim Denig for discussions and a careful reading of the manuscript. 
Special thanks also to Suzy Vascotto for careful proof-reading the manuscript.

\appendix

\section{The implementation of the final state emission}
                                         
 Final state emission of one photon and the final-initial state 
 interference terms are implemented in the present program in lowest order
  by means of
 the helicity amplitude method for both \(\pi^+ \pi^- \gamma\)
 and \(\mu^+ \mu^- \gamma\) final states. The notation is 
 the same as in \cite{RCKS} and will not be repeated.
 The pion-photon interaction is adopted
 from scalar electrodynamics.

 The helicity amplitudes describing the initial emission read

\begin{align}
 &M_\mathrm{ISR}(\lambda_{e^+},\lambda_{e^-},\lambda_1) =
 \frac{(4\pi\alpha)}{Q^2} \biggl\{ \non \\ &
 v_I^{\dagger}(p_1,\lambda_{e^+})
 A \ u_I(p_2,\lambda_{e^-})+v_{II}^{\dagger}(p_1,\lambda_{e^+})
 B \ u_{II}(p_2,\lambda_{e^-})\biggl\} \ ,
\end{align}

\noindent
where 
\begin{align}
 A &= 
\frac{\left(\varepsilon^*(k_1,\lambda_1)^- k_1^+ 
 - 2\varepsilon^*(k_1,\lambda_1)\cdot p_1\right)J^-}
 {2 k_1 \cdot p_1}\non \\ &
 +  \frac{J^-\left(2\varepsilon^*(k_1,\lambda_1)\cdot p_2 
 -  k_1^+ \varepsilon^*(k_1,\lambda_1)^-\right)}{2 k_1 \cdot p_2}
\end{align}

\noindent
and
\begin{align}
 B &= 
\frac{\left(\varepsilon^*(k_1,\lambda_1)^+ k_1^- 
 - 2\varepsilon^*(k_1,\lambda_1)\cdot p_1\right)J^+}
 {2 k_1 \cdot p_1}\non \\ &
 +  \frac{J^+\left(2\varepsilon^*(k_1,\lambda_1)\cdot p_2 
 -  k_1^- \varepsilon^*(k_1,\lambda_1)^+\right)}{2 k_1 \cdot p_2} \ .
\end{align}

The current \(J^\mu\) for \(\pi^+ \pi^-\) in the final state reads

 \begin{equation}
J^{\mu}_{2\pi} = i e F_{2\pi}(Q^2) \; (q_{\pi^+}-q_{\pi^-})^{\mu}~,
\label{hc}
\end{equation}

\noindent
while for  \(\mu^+ \mu^-\) in the final state it is given by

\begin{eqnarray}
 J^\mu_{2\mu}(\lambda_{\mu^+},\lambda_{\mu^-})
  = i e \bar u(q_1,\lambda_{\mu^-})\gamma^\mu 
 v(q_2,\lambda_{\mu^+}) \ \ .
\end{eqnarray}

The part of the amplitude that comes from the final state emission can
 be written as

\begin{align}
 &M_\mathrm{FSR}(\lambda_{e^+},\lambda_{e^-},\lambda_1) =
 \frac{(4\pi\alpha)}{s} \biggl\{ \non \\ &
 v_I^{\dagger}(p_1,\lambda_{e^+})
 D^- u_I(p_2,\lambda_{e^-})+v_{II}^{\dagger}(p_1,\lambda_{e^+})
 D^+ u_{II}(p_2,\lambda_{e^-})\biggl\} \ ,
\end{align}

\noindent
where the four-vector \(D^{\mu}\) reads

\begin{align}
&D^{\mu}(\lambda_1) = i e F_{2\pi}(s)\biggl\{
 \left(q_1+k_1-q_2\right)^{\mu} 
 \frac{q_1\cdot\varepsilon^*(k_1,\lambda_1)}{q_1\cdot k_1}\non \\ &
 +\left(q_2+k_1-q_1\right)^{\mu} 
 \frac{q_2\cdot\varepsilon^*(k_1,\lambda_1)}{q_2\cdot k_1}
 -2\varepsilon^{*\mu}(k_1,\lambda_1)\biggr\}  \ ,
\end{align}

\noindent
for \(\pi^+ \pi^-\) in the final state, while for  \(\mu^+ \mu^-\)
 in the final state it is

\begin{align}
&D^{\mu}(\lambda_1,\lambda_{\mu^+},\lambda_{\mu^-}) = i e\biggl\{\non \\ &
 u_I^{\dagger}(q_2,\lambda_{\mu^-})
 \tilde A^{\mu} v_I(q_1,\lambda_{\mu^+})
 +u_{II}^{\dagger}(q_2,\lambda_{\mu^-})
 \tilde B^{\mu} v_{II}(q_1,\lambda_{\mu^+})\biggl\} \ ,
\end{align}

\noindent
with

\begin{align}
\tilde A^{\mu} &= 
\frac{\left(\varepsilon^*(k_1,\lambda_1)^- k_1^+ 
 + 2\varepsilon^*(k_1,\lambda_1)\cdot q_2\right)\sigma^{\mu-}}
 {2 k_1 \cdot q_2}\non \\ &
 -  \frac{\sigma^{\mu-}\left(2\varepsilon^*(k_1,\lambda_1)\cdot q_1 
 +  k_1^+ \varepsilon^*(k_1,\lambda_1)^-\right)}{2 k_1 \cdot q_1} \ ,
\end{align}

\noindent
and

\begin{align}
\tilde B^{\mu} &= 
\frac{\left(\varepsilon^*(k_1,\lambda_1)^+ k_1^- 
 + 2\varepsilon^*(k_1,\lambda_1)\cdot q_2\right)\sigma^{\mu+}}
 {2 k_1 \cdot q_2}\non \\ &
 -  \frac{\sigma^{\mu+}\left(2\varepsilon^*(k_1,\lambda_1)\cdot q_1 
 +  k_1^- \varepsilon^*(k_1,\lambda_1)^+\right)}{2 k_1 \cdot q_1} \ .
\end{align}

The FSR matrix element squared and the FSR--ISR interference for pions
 in the final state agrees numerically
  (15 digits) with the code of EVA \cite{Binner:1999bt}, if
 non-leading mass terms \(\sim m_e^2\) missing in EVA are added. The largest
 relative change of the matrix element squared due to those missing terms
 is however as small as \(10^{-6}\). The sum over polarisations
 of the squared matrix element for muon final states is numerically
 identical to the result obtained by means of the trace method using
 FORM \cite{FORM}.
 For both final states 
 the external gauge invariance was checked numerically, while one can see
 at a glance that the above analytical formulae have that property. 
 
 To analyse the contribution from FSR
the program can be 
  run in three different options: initial state
 radiation only, initial state radiation plus final state radiation
 without interference
 and complete result with interference terms. In the last two cases
 it is necessary to change the generation of the phase space to
 absorb final state emission peaks. For these two options
 we use three channels to absorb the peaks in \(Q^2\) and in pion (muon)
 angular distributions. For the muon case we use the approximant

\begin{align}
 &f_{1}(q^2,\cos(\theta_{\mu})) = \frac{1}{1-q^2}+\frac{1}{q^2}\non \\ &
  +\frac{1}{1-q^2}\left(\frac{1}{1-v(m_\mu)\cos(\theta_{\mu})}
         +\frac{1}{1+v(m_\mu)\cos(\theta_{\mu})}\right) \ ,
\label{a1}
 \end{align}
 \noindent
where
\begin{align}
 v(m) = \sqrt{1-\frac{4m^2}{s}} \ \ ,  \ \ q^2 = \frac{Q^2}{s} \ \ , 
 \end{align}

For the pion case we use

\begin{align}
 &f_{2}(q^2,\cos(\theta_{\pi})) = \frac{1}{1-q^2}
 +\frac{1}{(q^2-m_{\rho}^2/s)^2
 + \Gamma_{\rho}^2 m_{\rho}^2/s^2}\non \\ &
  +\frac{1}{1-q^2}\left(\frac{1}{1-v(m_\pi)\cos(\theta_{\pi})}
         +\frac{1}{1+v(m_\pi)\cos(\theta_{\pi})}\right) \ .
\label{a2}
 \end{align}
\noindent

 An appropriate change of variables allows for a smoothing
 of the aforementioned 
 peaks. The \(q^2\) and  \(\cos(\theta_{\mu(\pi)})\)
 are generated according to the functions
 \(f_{1(2)}(q^2,\cos(\theta_{\mu(\pi)}))\).
  These very simple approximants work well enough to allow
 for relatively fast event generation.

\section{The analytical formulae used in Section 2}

The formulae resulting from the analytical evaluation of the integration over 
 photon angles are adopted from Refs. \cite{Berends:1986yy,Berends:1988ab}
  (see also Section 2 for details). The contributions of the 
 virtual + soft corrections to the hadronic invariant mass (\(Q^2\))
 differential distribution are given by
\begin{align}
& Q^2 \frac{d\sigma}{dQ^2} = \frac{4\alpha^3}{3 s} R(Q^2)
\bigg\{ \frac{1+q^4}{1-q^2} (L-1) \non \\ & \times
\bigg( 1 + \frac{\alpha}{\pi} \; \bigg[ \log(4 w^2) (L-1) 
+ \frac{3}{2} L - 2 + \frac{\pi^2}{3} \bigg] \bigg) \non \\ &
+ \frac{\alpha}{\pi} \; \bigg[- \frac{1+q^4}{2(1-q^2)} \; \log(q^2) \; L^2 
\non \\ &
+ \bigg\{ \frac{1+q^4}{1-q^2} \bigg( \Li_2(1-q^2) + \log(q^2)\log(1-q^2) 
\non \\ & \quad - \frac{\log^2(q^2)}{2} 
+ \frac{5}{2} \log(q^2) \bigg) - (1-q^2) \log(q^2)
 +\frac{q^2}{2} \bigg\} L \non \\ &
+ \frac{1+q^4}{1-q^2} \bigg( S_{1,2}(1-q^2) 
+ \bigg[ \log(q^2) - \frac{3}{2} \bigg] \Li_2(1-q^2) \non \\ & \quad
+ \bigg[ \log(q^2)\log(1-q^2) - \frac{\log^2(q^2)}{3}+ \log(q^2) 
\non \\ & \quad - 3\log(1-q^2)  - 8 \bigg] 
\frac{\log(q^2)}{2} \bigg) \non \\ &
+ (1+q^2) \bigg( 2 \Li_3(1-q^2) - S_{1,2}(1-q^2) \non \\ & \quad
- \log(1-q^2) \Li_2(1-q^2) + \frac{\log^2(q^2)}{4} \bigg) \non \\ &
+ \frac{1-7q^2}{2} \bigg(\Li_2(1-q^2)+\log(q^2)\log(1-q^2) \bigg) \non \\ &
- \frac{1-5q^2}{4} \bigg( \log^2(1-q^2)+\frac{2\pi^2}{3} \bigg) \non \\ &
+ \frac{3-2q^2}{2} \log(1-q^2)+\frac{7-5q^2}{2}\log(q^2)-1
\bigg] \bigg\}~,
\label{eq:virtualsoft}
\end{align}
While the emission of
two hard photons, i.e. both photons with energy larger than $w \sqrt{s}$,
contributes as
\begin{align}
& Q^2 \frac{d\sigma}{dQ^2} = \frac{4\alpha^3}{3 s} R(Q^2) \frac{\alpha}{\pi}
\non \\ & \times \bigg[ \frac{1+q^4}{1-q^2} 
\bigg\{ 2\log(\frac{1-q^2}{2 w}) -\frac{\log(q^2)}{2} \bigg\} (L-1)^2
\non \\ & + 
\bigg\{ -(1-q^2) + (1+q^2) \frac{\log(q^2)}{2} \bigg\} L^2 \non \\ & +
\bigg\{ \frac{7}{2}(1-q^2)-q^2\log(q^2)+(1+q^2)\frac{\log^2(q^2)}{4}\bigg\} L
\non \\ & +
\frac{1+q^4}{1-q^2} \bigg( -S_{1,2}(1-q^2)-\frac{\log(q^2)}{2} \Li_2(1-q^2)
\non \\ &
-\frac{3\log^2(q^2)}{2} + \big(\frac{\pi^2}{6}+\frac{5}{3}\big) \log(q^2)
\bigg) \non \\ \non
\end{align}
\begin{align}
& -
(1+q^2) \bigg( \frac{\Li_3(1-q^2)}{2}+S_{1,2}(1-q^2) \bigg) 
-\frac{\pi^2 q^2}{9} \non \\ & -
\bigg( \frac{1}{2}+\frac{2q^2}{3} \bigg) \Li_2(1-q^2)
 -(10-25 q^2) \frac{\log(q^2)}{6}  \non \\ &
+ \bigg( \frac{2}{(1-q^2)^2} 
-\frac{1}{4}-\frac{7q^2}{3}\bigg) \log^2(q^2) \non \\ & +
\frac{1-q^2}{2}-\frac{2}{3} \frac{q^2}{1-q^2}
\bigg(1+\frac{\log(q^2)}{1-q^2}\bigg)^2 \bigg]~,
\label{eq:twohard}
\end{align}
with $L=\log(s/m_e^2)$ and $S_{1,2}$ the Nielsen's generalised 
polylogarithm function
\begin{equation}
S_{n,p}(z) = \frac{(-1)^{n+p-1}}{(n-1)! \  p!} 
\int_0^1 \log^{n-1}(t) \frac{\log^p(1-zt)}{t} dt~,
\end{equation}
 $\Li_n(z)=S_{n-1,1}(z)$ being the polylogarithms
\begin{equation}
\Li_n(z) = \sum_{k=1}^\infty \frac{z^k}{k^n}~, \qquad |z|<1~,
\end{equation}
and
\begin{align}
S_{1,2}(1-z) &= \frac{1}{2}\log^2(z)\log(1-z)+\zeta(3) \non \\
&+ \log(z) \Li_2(z)-\Li_3(z)~.
\end{align}

The function \(R(Q^2)\) is related to the hadronic current
  \(J^{em}\) through

\begin{eqnarray}
 &&{\kern-60pt}
 \int \ J^{em}_\mu (J^{em}_\nu)^*  \ \ d\bar\Phi_n(Q;q_1,\dots,q_n) =
 \nonumber \\
 &&{\kern+20pt}
 \frac{1}{6\pi} \left(Q_{\mu}Q_{\nu}-g_{\mu\nu}Q^2\right) \ R(Q^2) \ \ ,
\label{rr}
\end{eqnarray}                                                          
        
 \noindent
where \(d\bar\Phi_n(Q;q_1,\dots,q_n)\) denotes the \(n\)-body phase space
 with all statistical factors coming from the hadronic final state included. 
     
The ratio 
 \(R(Q^2) =\) \(\sigma(e^+e^-\rightarrow {\mathrm{ hadrons}})/\sigma_{point}\)
  for
  hadrons = \(\pi^+\pi^-\) is equal to

\begin{eqnarray}
 R(Q^2) = |F(Q^2)|^2 \frac{\beta_\pi^3}{4} \ .
 \label{rr1}
\end{eqnarray}



\begin{thebibliography}{99}

\bibitem{Brown:2001mg}
H.~N.~Brown {\it et al.} [Muon $g-2$ Collaboration],
Phys. Rev. Lett. {\bf 86} (2001) 2227 [hep-ex/0102017]; G.W.Bennett
{\it et al.} [Muon $g-2$ Collaboration], Phys. Rev. Lett.
{\bf 89} (2002) 101804, Erratum, ibid. {\bf 89} (2002) 129903,
[hep-ex/0208001].                         

\bibitem{Jegetc.}
F.~Jegerlehner,
hep-ph/0104304. 

\bibitem{HMNT02}
K. Hagiwara, A.D. Martin, Daisuke Nomura and T. Teubner, 
hep-ph/0209187.

\bibitem{Davier:2002dy}
M.~Davier, S.~Eidelman, A.~H\"ocker and Z.~Zhang,
hep-ph/0208177.  

\bibitem{BES}
W.B. Yan, W.G. Li and Z.G. Zhao [BES Collaboration], hep-ex/021000.

\bibitem{CMD2}
R.R. Akhmetshin {\it et al.}, Phys. Lett. B {\bf 527} (2002) 161
[hep-ex/0112031].
 
\bibitem{Binner:1999bt}
S.~Binner, J.~H.~K\"uhn and K.~Melnikov,
Phys.\ Lett.\ B {\bf 459} (1999) 279
[hep-ph/9902399].
 
\bibitem{Melnikov:2000gs}
K.~Melnikov, F.~Nguyen, B.~Valeriani and G.~Venanzoni,
Phys.\ Lett.\  {\bf B477} (2000) 114
[hep-ph/0001064].

\bibitem{Czyz:2000wh}
H.~Czy\.z and J.~H.~K\"uhn,
Eur.\ Phys.\ J.\ C {\bf 18} (2001) 497
[hep-ph/0008262].
 
\bibitem{Spagnolo:1999mt}
S.~Spagnolo,
Eur.\ Phys.\ J.\ C {\bf 6} (1999) 637.

\bibitem{CCR}M. Caffo, H. Czy{\.z} and
E. Remiddi, Nuovo Cim. {\bf 110A} (1997) 515 [hep-ph/9704443];
Phys. Lett. B {\bf 327}(1994)369.                     

\bibitem{RCKS}
G. Rodrigo, H. Czy{\.z}, J.H. K{\"u}hn and M. Szopa,
Eur.\ Phys.\ J.\ C {\bf 24} (2002) 71
[hep-ph/0112184].

\bibitem{Rodrigo:2001jr}
G.~Rodrigo, A.~Gehrmann-De Ridder, M.~Guilleaume and J.~H.~K\"uhn,
Eur.\ Phys.\ J.\ C {\bf 22} (2001) 81
[hep-ph/0106132].                                                              

\bibitem{RK02}
G. Rodrigo and J.H. K{\"u}hn,
Eur.\ Phys.\ J.\ C {\bf 25} (2002) 215
[hep-ph/0204283].

\bibitem{Aloisio:2001xq}
A.~Aloisio {\it et al.}  [KLOE Collaboration],
hep-ex/0107023.
 
\bibitem{Denig:2001ra}
A.~Denig {\it et al.}  [KLOE Collaboration],
eConf {\bf C010430} (2001) T07
[hep-ex/0106100].
 
\bibitem{Adinolfi:2000fv}
M.~Adinolfi {\it et al.}  [KLOE Collaboration],
hep-ex/0006036.

\bibitem{Barbara:Morion}
B. Valeriani {\it et al.} [KLOE Collaboration], 
hep-ex/0205046. 

\bibitem{Venanzoni:2002}
G. Venanzoni {\it et al.} [KLOE Collaboration], hep-ex/0210013;
hep-ex/0211005.

\bibitem{Achim:radcor02}
A.~Denig {\it et al.}  [KLOE Collaboration],
hep-ex/0211024.
 
\bibitem{babar}
E.~P.~Solodov  [BABAR collaboration],
eConf {\bf C010430} (2001) T03 [hep-ex/0107027]. 

\bibitem{Berger:2002mg}
N.~Berger,
hep-ex/0209062.

\bibitem{Kuhn:1990ad}
J.~H.~K\"uhn and A.~Santamaria,
Z.\ Phys.\ C {\bf 48} (1990) 445.
 
\bibitem{Decker:1994af}
R.~Decker, M.~Finkemeier, P.~Heiliger and H.~H.~Jonsson,
Z.\ Phys.\ C {\bf 70} (1996) 247
[hep-ph/9410260].
 
\bibitem{Ecker:2002cw}
G.~Ecker and R.~Unterdorfer,
Eur. Phys. J. C {\bf 24} (2002) 535 [hep-ph/0203075].

\bibitem{Kolodziej:1991pk}
K.~Ko\l odziej and M.~Zra\l ek,
Phys.\ Rev.\ D {\bf 43} (1991) 3619.

\bibitem{Jegerlehner:2000wu}
F.~Jegerlehner and K.~Ko\l odziej,
Eur.\ Phys.\ J.\ C {\bf 12} (2000) 77
[hep-ph/9907229].

\bibitem{Rodrigo:2002hk}
G.~Rodrigo, H.~Czy\.z and J.~H.~K\"uhn,
 hep-ph/0205097; hep-ph/0210287; hep-ph/0211186.

\bibitem{Berends:1986yy}
F.~A.~Berends, G.~J.~Burgers and W.~L.~van Neerven,
Phys.\ Lett.\ B {\bf 177} (1986) 191.
 
\bibitem{Berends:1988ab}
F.~A.~Berends, W.~L.~van Neerven and G.~J.~Burgers,
Nucl.\ Phys.\ B {\bf 297} (1988) 429;
Erratum, ibid. B {\bf 304} (1988) 921.                            

\bibitem{Rodrigo:2001cc}
G.~Rodrigo,
Acta Phys.\ Polon.\ B {\bf 32} (2001) 3833 [hep-ph/0111151].

\bibitem{CDKMV2000} 
G. Cataldi, A. Denig, W. Kluge, S. Muller and G. Venanzoni,
Frascati Physics Series (2000) 569.

\bibitem{K2} J.H.~K\"uhn, Nucl. Phys. B (Proc. Suppl.) {\bf 76} (1999) 21.  

\bibitem{FORM} J. A. M. Vermaseren, {\it Symbolic Manipulations with FORM},
Computer Algebra Nederland, Amsterdam, 1991.
\end{thebibliography}
\end{document}